\def\bq{\begin{equation}}
\def\eq{\end{equation}}
\def\bqa{\begin{eqnarray}}
\def\eqa{\end{eqnarray}}
\def\bqb{\begin{eqnarray*}}
\def\eqb{\end{eqnarray*}}
\documentstyle[epsf,12pt]{article}
\def\eg{{\it e.g.\/}}

\global\nulldelimiterspace = 0pt

\def\Bsl{\hbox{/\kern-.6700em$B$}} 
\def\Dsl{\hbox{/\kern-.6700em$D$}} 
\def\Wsl{\hbox{/\kern-.6700em$W$}} 

\def\roughly#1{\mathrel{\raise.3ex
    \hbox{$#1$\kern-.75em\lower1ex\hbox{$\sim$}}}}

\def\gsim{\roughly>}

\def\ol#1{\overline{#1}}

\def\L{ {\cal L }}
\def\O{ {\cal O }}

\def\mwd{M_W^2}

\def\lw{\lambda_W}

\def\mh2{m^2_H}

\begin{document}
\pagenumbering{arabic}
\thispagestyle{empty}
\begin{flushleft}
 PM/96-08 \\ THES-TP 96/02 \\
February 1996 \end{flushleft}
\vspace {0.8cm} 

\begin{center}
{\Large\bf Residual New Physics Effects in $e^+e^-$ and
$\gamma\gamma$ Collisions at NLC}
\footnote{Partially supported by the EC contract CHRX-CT94-0579.}
 \vspace{0.8cm}  \\
{\large G.J. Gounaris$^{a}$, J.Layssac$^b$, J.E. Paschalis$^a$,  
F.M. Renard$^b$ \\
and N.D. Vlachos$^{a}$} 
\vspace {0.1cm}  \\

$^a$Department of Theoretical Physics, University of Thessaloniki,\\%
Gr-54006, Thessaloniki, Greece,\\\vspace{0.2cm}
$^b$Physique
Math\'{e}matique et Th\'{e}orique, CNRS-URA 768,\\Universit\'{e} de
Montpellier II,  F-34095 Montpellier Cedex 5.\\
\vspace{1.5cm}

{\bf Contribution to the NLC workshop}\\

\vspace{0.5cm}

 {\bf Abstract}
\end{center}
\noindent
We report on a set of studies concerning the description 
of New Physics (NP) effects characterized by a scale 
much higher than the electroweak
scale. We concentrate on the residual effects described by an effective
lagrangian involving only bosonic fields and restrict to the case
that the Higgs field is described linearly, so that the Higgs
particle exists. It has already been
emphasized that this lagrangian may be represented by
eight $dim=6$ operators; namely the blind CP-conserving
operators 
$\O_W$, $\O_{B\Phi}$, $\O_{W\Phi}$,
$\O_{UB}$, $\O_{UW}$, the
''superblind'' ${\cal O}_{\Phi 2}$ and the two 
CP-violating ones $\ol{\O}_{UB}$
and $\ol{\O}_{UW}$.
To each operator and for any given value of its coupling,
we associate an unambiguous NP scale determined  by
the energy where  unitarity is saturated.
Our study of the possible NP tests realizable at a future 
linear $e^+e^-$ collider concentrates on $(e^+e^-\to
HZ ~,~ H\gamma)$, and on $\gamma\gamma$ collisions producing 
boson pairs ($\gamma\gamma \to W^+W^-$, $ZZ$,
$Z\gamma$, $\gamma\gamma$, $HH$),  or a single Higgs ($\gamma\gamma 
\to H$). The sensitivities to the
various operators involved in the effective lagrangian
are estimated. We find that the testable NP scales vary from a few
tens of TeV in the boson pair processes,
up to 65 TeV in single H production.
Ways to disentangle the various operators are also proposed.
In this respect a comparison of
the NP effects in the various boson pair processes, like \eg\@
a comparison of $e^+e^-\to
HZ$ versus $e^+e^-\to H\gamma$,
should be very fruitful. This is applied also to the study of 
the Higgs branching ratios, and especially the ratio
$\Gamma(H\to\gamma\gamma)/\Gamma (H\to\gamma Z)$.

\clearpage

\section{Introduction}

The search for manifestations of New Physics (NP) is an important 
part of the
program of future high energy colliders.
 If it happens that all
new particles are too heavy to be directly produced at these colliders,
then the only way NP could manifest itself, is through residual 
interactions
affecting the particles already present in SM. The possibility for such
residual NP effects involving the interactions of the gauge 
bosons with the
leptons and to some extent the light quarks, has already been 
essentially
excluded by LEP1. Thus, the self-interactions among the
gauge and Higgs bosons and possibly also the heavy quarks 
constitute the most probable space for some NP.
On the other hand 
the scalar sector is the most mysterious part of the Standard Model
 (SM) of
the electroweak interactions and  is
the favorite place for generating New
Physics (NP) manifestations.\par

 High energy $e^+e^-$ linear colliders will offer many
possibilities to test the sector of the gauge
boson and Higgs interactions with a high
accuracy, firstly in $e^+e^-$ collisions and secondly through
$\gamma\gamma$ collisions.
The most famous process is $e^+e^- \to W^+W^-$. This has
been carefully studied and it has been shown that indeed the 3-gauge
boson vertices $\gamma W^+W^-$ and $Z W^+W^-$ can be very
accurately constrained through it.
If the Higgs boson is not too heavy further processes to consider are
$e^{-}e^{+}\to ZH,\,\gamma H$.
In SM, the process 
$e^{-}e^{+}\to Z\to ZH$ is
allowed at tree level,
while $e^{-}e^{+}\to \gamma H$ is only possible at 1-loop level 
and is therefore a dedicated
place to look for anomalous Higgs-gauge boson interactions. \par

A new facility offered by high energy linear $e^+e^-$ colliders 
is the
realization of $\gamma\gamma$ collisions with intense high energy
photon beams through the laser backscattering
method. This will give access to
a new kind of processes namely boson-boson scattering.
Five boson pair
production processes are thus accessible, 
$\gamma\gamma \to 
W^+W^-$, $ZZ$, $Z\gamma$, $\gamma\gamma$ and $HH$. Such 
processes are very interesting, since they are
sensitive not only to the 3-gauge boson vertices, but also to 
4-gauge boson as well as to
Higgs couplings. 
In addition, provided the Higgs particle will be
accessible at the future colliders, the $\gamma\gamma$ fusion
into a single Higgs boson will give direct access to the study
of this particle and thus test the scalar 
sector in a much deeper
way, particularly because the standard contribution 
to $\gamma\gamma\to H$
 only occurs at
1-loop.  With the high luminosities expected
at linear $e^+e^-$ colliders, thousands of Higgs bosons
should be produced. The 
sensitivity to anomalous $H\gamma\gamma$
couplings is therefore very strong. \par

Below we summarize the studies in 
\cite{hzVlachos, higpro2, ggVV2}
and show that they provide very sensitive tests of the
various possible forms of\footnote{Because of lack of space, 
the extensive litterature on
the subject is not given explicitely and 
can be found in \cite{hzVlachos, higpro2, ggVV2}.}
NP.

Assuming that the NP scale $\Lambda_{NP}$ is sufficiently large,
the effective NP Lagrangian should be satisfactorily 
described  in terms of 
$dim=6$ bosonic operators only.
There exist only seven $SU(2)\times U(1)$ gauge
invariant such operators called "blind",
which are not strongly constrained by existing LEP1 
experiments. Three of them

\bq
\O_W =  {1\over3!}\left( \overrightarrow{W}^{\ \ \nu}_\mu\times
  \overrightarrow{W}^{\ \ \lambda}_\nu \right) \cdot
  \overrightarrow{W}^{\ \ \mu}_\lambda \ \ , \ 
\eq
\bq 
\O_{W\Phi}  =  i\, (D_\mu \Phi)^\dagger \overrightarrow \tau
\cdot \overrightarrow W^{\mu \nu} (D_\nu \Phi) \ \ \ , \ \ \  
\O_{B\Phi}  =  i\, (D_\mu \Phi)^\dagger B^{\mu \nu} (D_\nu
\Phi)\ \ \   \
\eq
\noindent
induce anomalous triple gauge boson couplings, while
the remaining four
\bq
\O_{UW}  =  \frac{1}{v^2}\, (\Phi^\dagger \Phi - \frac{v^2}{2})
\, \overrightarrow W^{\mu\nu} \cdot \overrightarrow W_{\mu\nu} \ \ \
 , \ \ \
\O_{UB}  =  \frac{4}{v^2}~ (\Phi^\dagger \Phi -\frac{v^2}{2})
B^{\mu\nu} \
B_{\mu\nu} \ \ ,
\eq
\bq 
\ol{\O}_{UW}  =  \frac{1}{v^2}\, (\Phi^\dagger
\Phi)
\, \overrightarrow W^{\mu\nu} \cdot
\widetilde{\overrightarrow W}_{\mu\nu} \ \ \ , \ \
\ol{\O}_{UB}  =  \frac{4}{v^2}~ (\Phi^\dagger \Phi )
B^{\mu\nu} \
\widetilde{B}_{\mu\nu} 
\eq

\noindent
create  anomalous CP conserving and CP violating
Higgs couplings.

In addition, the ''superblind'' operator
\begin{equation}
{\cal O}_{\Phi 2}~=~4\partial _\mu (\Phi ^{\dagger }\Phi )\partial ^\mu
(\Phi ^{\dagger }\Phi )\ \ \ \ ,\ 
\end{equation}
 which is insensitive to ~LEP1 physics even at the 1-loop level, 
induces a wave
function renormalization to the Higgs field that may be observable
in  $e^{-}e^{+}\to ZH$.

The effective Lagrangian describing the NP induced by these
operators is given by
\bqa
\L_{NP} & = & \lw \frac{g}{\mwd}\O_W +
{f_B g\prime \over {2M^2_W}}\O_{B\Phi} +{f_W g \over
{2M^2_W}}\O_{W\Phi}\ \ + \nonumber \\
\null & \null &
d\ \O_{UW} + {d_B\over4}\ \O_{UB} +\ol{d}\ \ol{\O}_{UW} +
{\ol{d}_B\over4}\ \ol{O}_{UB} 
 +\frac{f_{\Phi 2}}{v^2}{\cal O}_{\Phi 2}\ \ \ \
\ \ ,  \
\eqa
which  defines the various couplings.  Studying  unitarity,
we have established relations
between these coupling constants and the corresponding  NP 
scales $\Lambda_{NP}$ (defined as
the energy value at which unitarity is saturated). This  allows us to
express the observability limits directly in terms of
 $\Lambda_{NP}$.\par

We now study how the aforementioned processes accessible at NLC react
to each of these operators and what is the observability limit for the
associated NP scale. We also give  ways to disentangle the effects
of the various operators.

\section{Analysis of $e^-e^+\to Z_{(f\bar f)} H$ 
and $e^-e^+\to \gamma H$%
.}

Using the NP lagrangian given by (6),
we have computed in \cite{hzVlachos} the helicity 
amplitude for $e ^{-}e^{+}\to Z H$.  
They depend on the SM couplings and on the combinations of NP couplings 
$d_{\gamma Z}=s_Wc_W(d-d_B)$ , $d_{ZZ}=dc_W^2+d_Bs_W^2$ , $d_{\gamma
\gamma }=ds_W^2+d_Bc_W^2$ ,
as well as the corresponding $CP$-violating ones for 
$\ol{d},~\ol{d}_B$  defined
in (6).\par
  
The corresponding total cross section as a function of the
$e^{-}e^{+}$
energy is shown in Fig.1 assuming $m_H=80GeV$. In this
figure, we present results for various values 
of the $d$ and $d_B$ couplings,
indicating also the  NP scale they correspond to. Identical
results are of course also obtained for the same values of the 
$CP$-violating couplings $\overline{d}$ 
and $\overline{d}_B$ respectively.
The $HZ$ angular distribution depends on $f_{\Phi 2}$ through
 the Higgs field renormalization factor $Z_H$, while its 
dependence on $\O_{UW}$ and  $\O_{UB}$ is mainly given by 
 the combination 
$d_{ZZ}$ (see Fig.2).


\[
\epsfxsize=7cm
\hspace*{1cm}\epsffile[97 0 576 502]{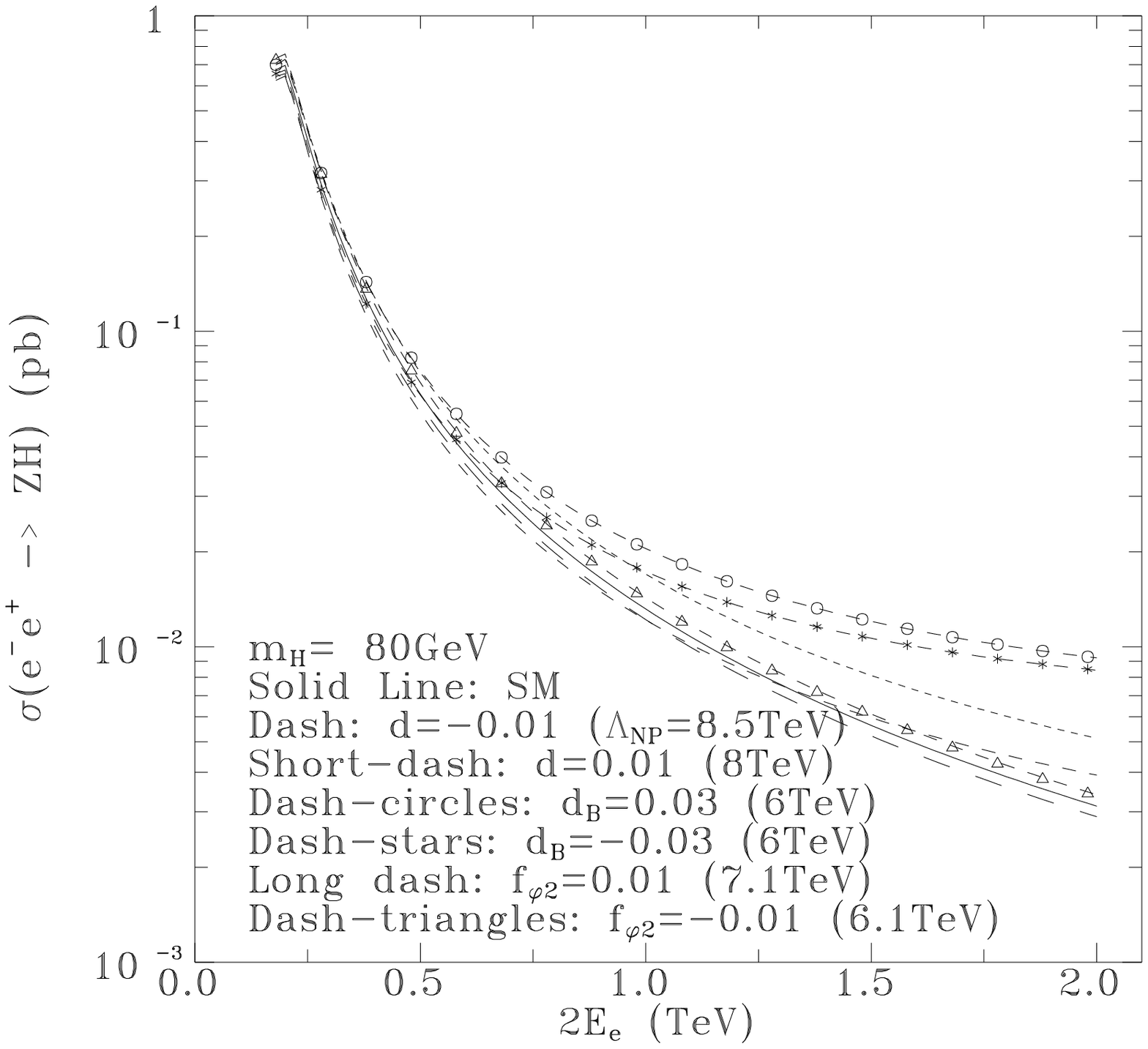}\ \ \ \ \ 
\epsfxsize=7cm
\epsffile[97 0 576 502]{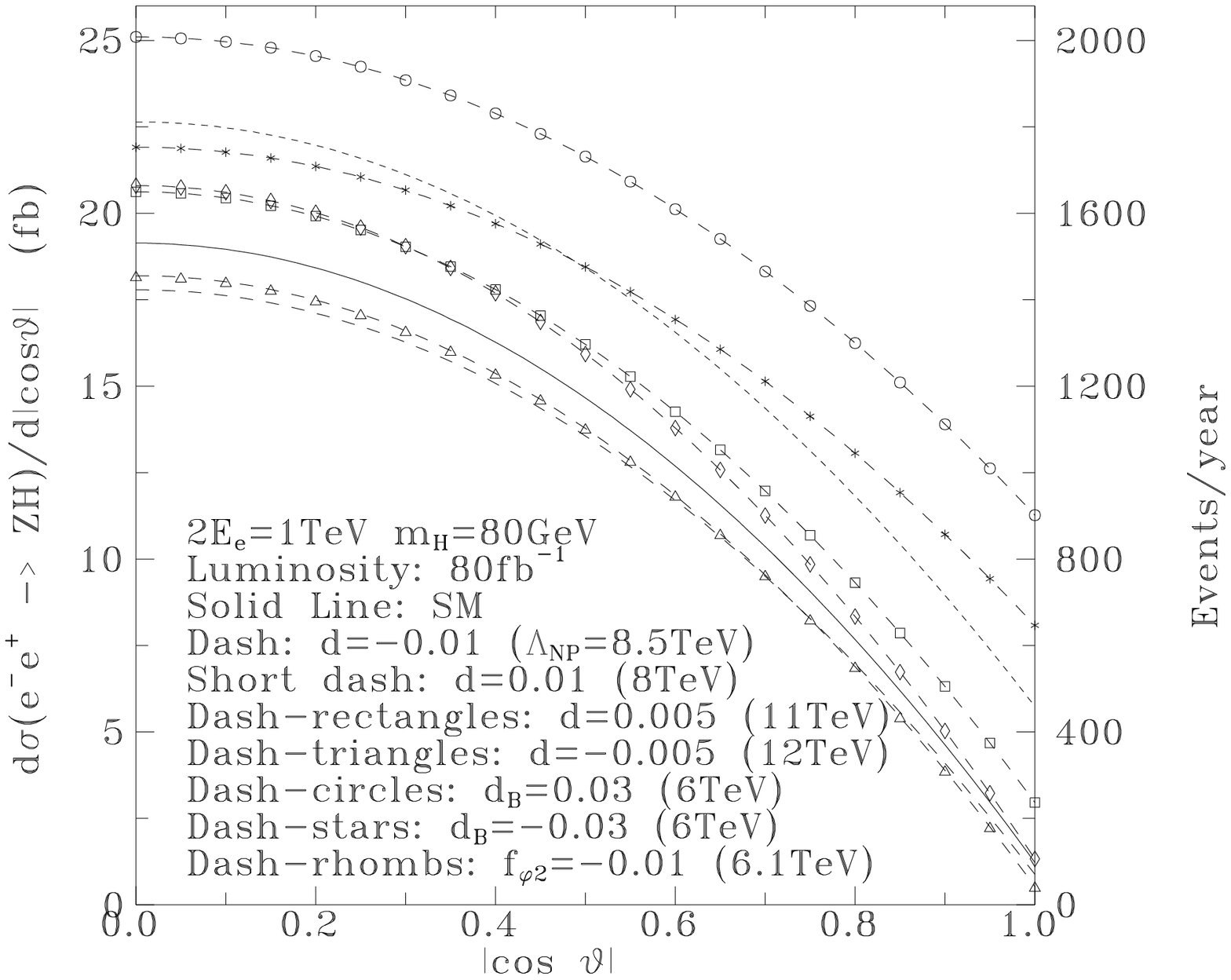}
\]
\vspace{-1.5cm}
\null\\ \vspace{0.5cm}
\centerline{\bf \hfill \  Fig 1 \hfill Fig 2  \hfill}

Thus at 1 $TeV$, with 80 $fb^{-1}$ per year, the SM rate
gives one thousand 
events per year. One can then expect a measurement 
of the cross section with
a 3\% accuracy. The implied sensitivities to the various couplings
are $|f_{\phi 2}|\simeq 0.004$, $|d|\simeq 0.005$ 
and $|d_B|\simeq 0.015$,
corresponding to NP scales of 10, 11 and 9 $TeV$ respectively;
compare Fig.1 and Fig.2. Moreover, with such a number of events, 
informations from the fermionic decay distribution of the $Z$ can 
be used in order to disentangle the various $d$-type
couplings.\par

We then compute the $Z$  density matrix in 
the helicity basis and,
assuming that the decay $Z\to f\bar f$ is standard, the
angular distribution for the $f\bar f$ system in the $Z$ rest frame. 
In \cite{hzVlachos} it was shown that the $\phi_f$ 
azimuthal distribution  allows to
define four asymmetries $A_{13}$, $A_{12}$, $A_{14}$, $A_{8}$,
respectively associated to the $\sin2\phi_f$, $\cos2\phi_f$,
$\sin\phi_f$, $\cos\phi_f$ dependences. The remarkable features are
that $A_{14}$, $A_{12}$ are mainly sensitive  on the combination 
$d_{\gamma Z}$ and on its 
$CP$-violating partner $\overline{d}_{\gamma Z}$. 
Thus, the illustrations made for $d$ ,($\overline{d}$)  apply also to
 $-d_B$, ($-\overline{d}_B$).
They are shown in Figs.3,4. These asymmetries depend
on the flavour of the fermion $f$ to which $Z$ decays and are mostly
interesting for the $Z\to b\bar b$ case. 
The asymmetry $A_{13}$ shown in Fig.5 does not depend
on the type of the fermion $f$ and it is sensitive to the $CP$-violating
combination $\overline{d}_{ZZ}$. Thus,~sensitivities to
 this
coupling at the percent level should be possible using this asymmetry.
 Note
that in this case the sensitivities to 
$\overline{d}_B$ and $\overline{d}$ are related by 
$\overline{d}_B\simeq \overline{d}c_W^2/s_W^2$.
 Finally the asymmetry $A_8$ shown in Fig.6,
 is also
mainly sensitive to $d_{ZZ}$ and independent of the flavour of the $f$
fermion. Consequently, the ~differential cross section for 
$e^{-}e^{+}\to ZH$, together
with the above asymmetries, provide considerable information for 
disentangling the aforementioned five relevant anomalous 
couplings at the percent level.
It follows that NP scales of the order of a few $TeV$ could be 
searched for this way.\par
\newpage

\[
\hspace*{1.5cm}\epsfxsize=7cm
\epsffile[97 0 576 502]{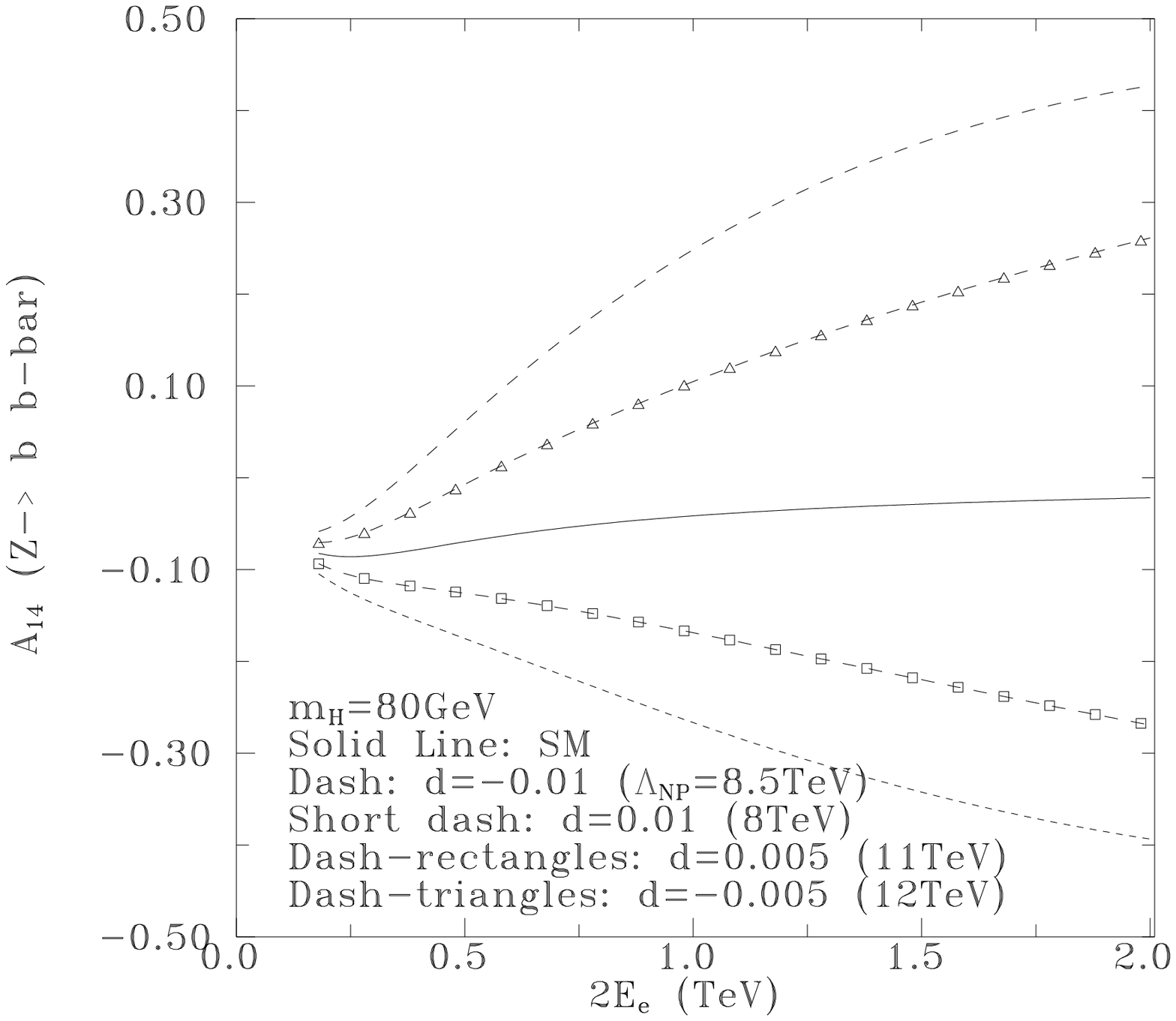}\ \ \ \ \ \ \ \ \
\epsfxsize=7cm
\epsffile[97 0 576 502]{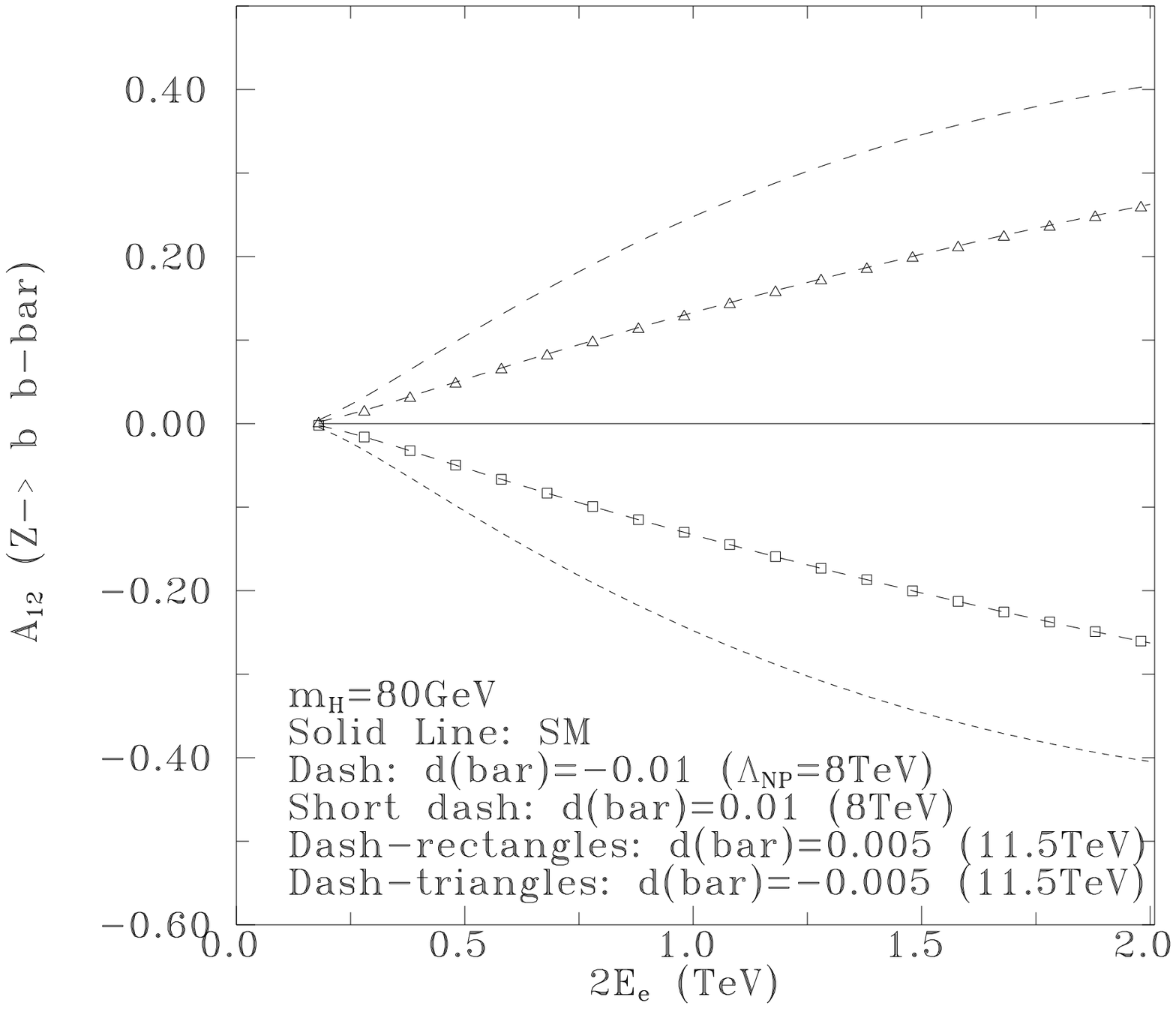}
\]
\vspace{-1.5cm}
\null\\ \vspace{0.5cm}
\centerline{\bf \hfill \  Fig 3 \hfill Fig 4  \hfill}
\[
\hspace*{1.5cm}\epsfxsize=7cm
\epsffile[97 0 576 502]{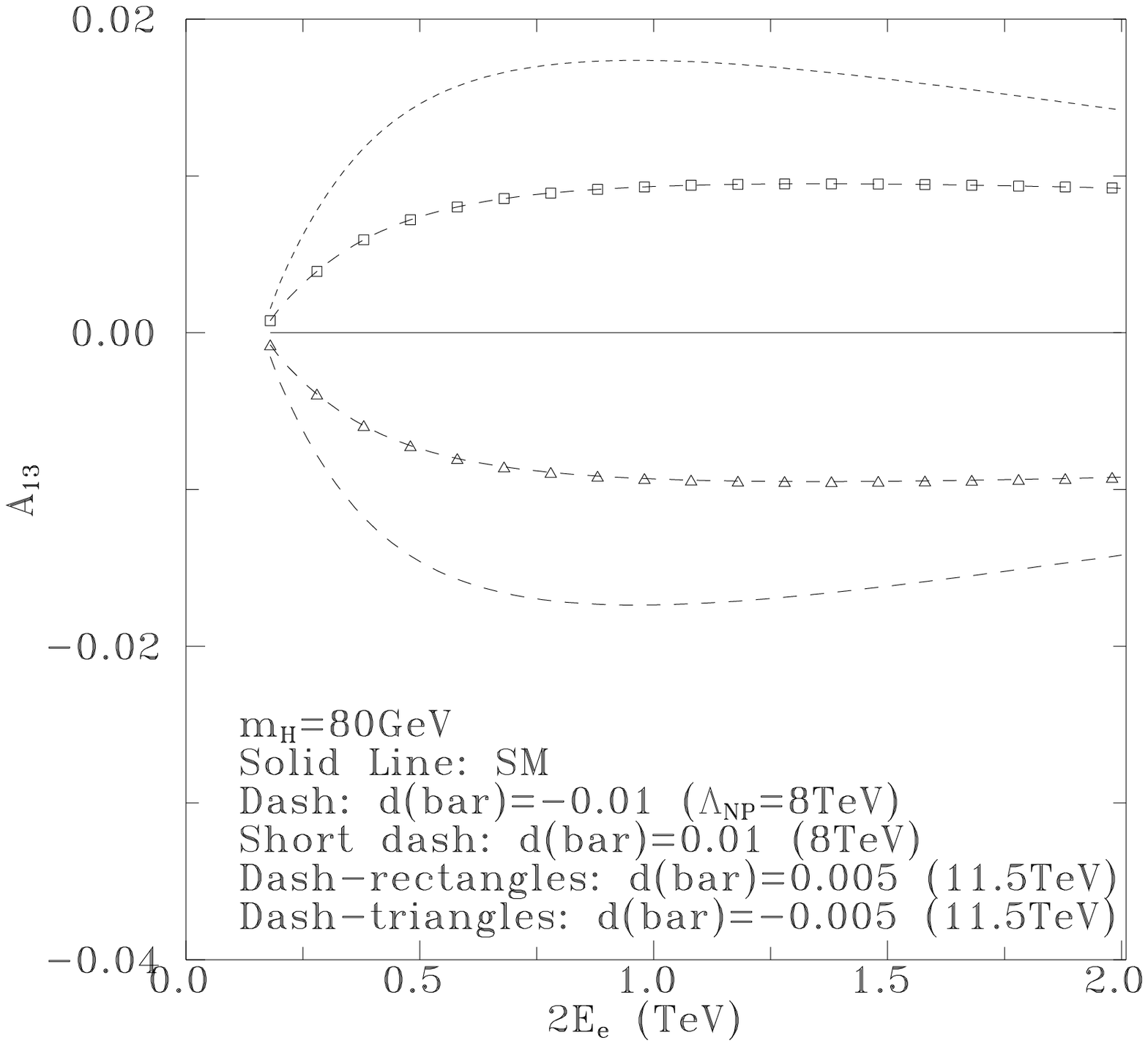}\ \ \ \ \ \ \ \ \
\epsfxsize=7cm
\epsffile[97 0 576 502]{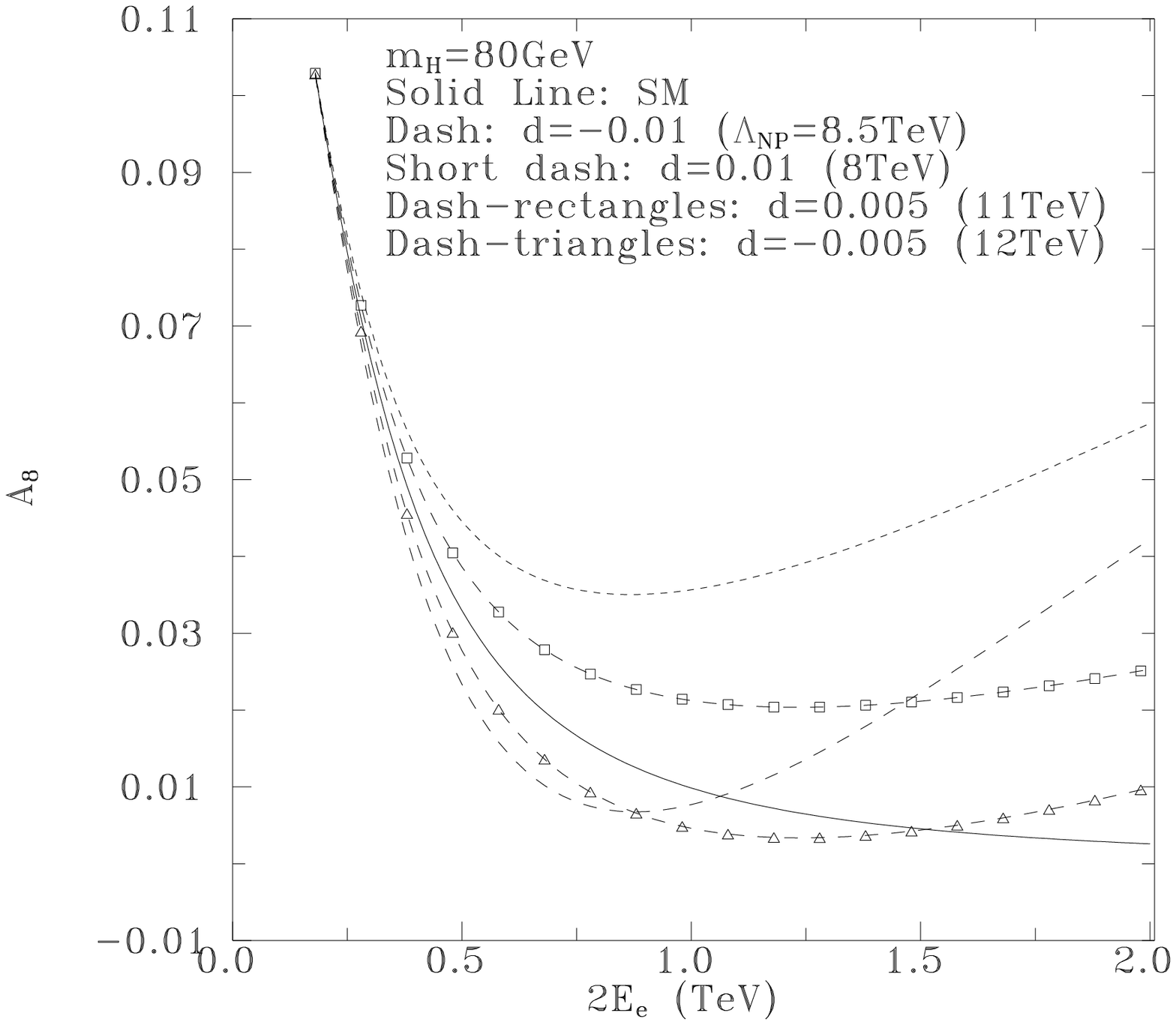}
\]
\vspace{-1.5cm}
\null\\ \vspace{0.5cm}
\centerline{\bf \hfill \  Fig 5 \hfill Fig 6  \hfill}

We next turn to the process
$e^{-}e^{+}\to \gamma H$. In this case 
\cite{hzVlachos} the
NP contribution is added to 
the 1-loop SM contribution.
We have again neglected
tree-level contributions quadratic in the anomalous couplings, and also
1-loop contributions linear in the anomalous couplings. This implies that no
contribution from ${\cal O}_{\Phi 2}$ should be included.
At SM the cross section is unobservably small, but it is very sensitive
to the four NP interactions and provides informations on combinations
of couplings that are different from those appearing in $e^+e^-\to
ZH$. The differential cross section could
become observable if non negligible anomalous interactions occur
. 
For a $1TeV$ NLC, Fig.7a,b show the sensitivity to $|d|\simeq
0.005$ and $|d_B|\simeq 0.0025$ corresponding to NP scales of 11 $TeV$ 
and 22 $TeV$ respectively. In these figures we have used for
illustration $m_H=80GeV$, but of course similar results would
have been expected for any $m_H \ll 1TeV $. 

\[
\epsfxsize=7cm
\hspace*{1.cm}\epsffile[97 0 576 502]{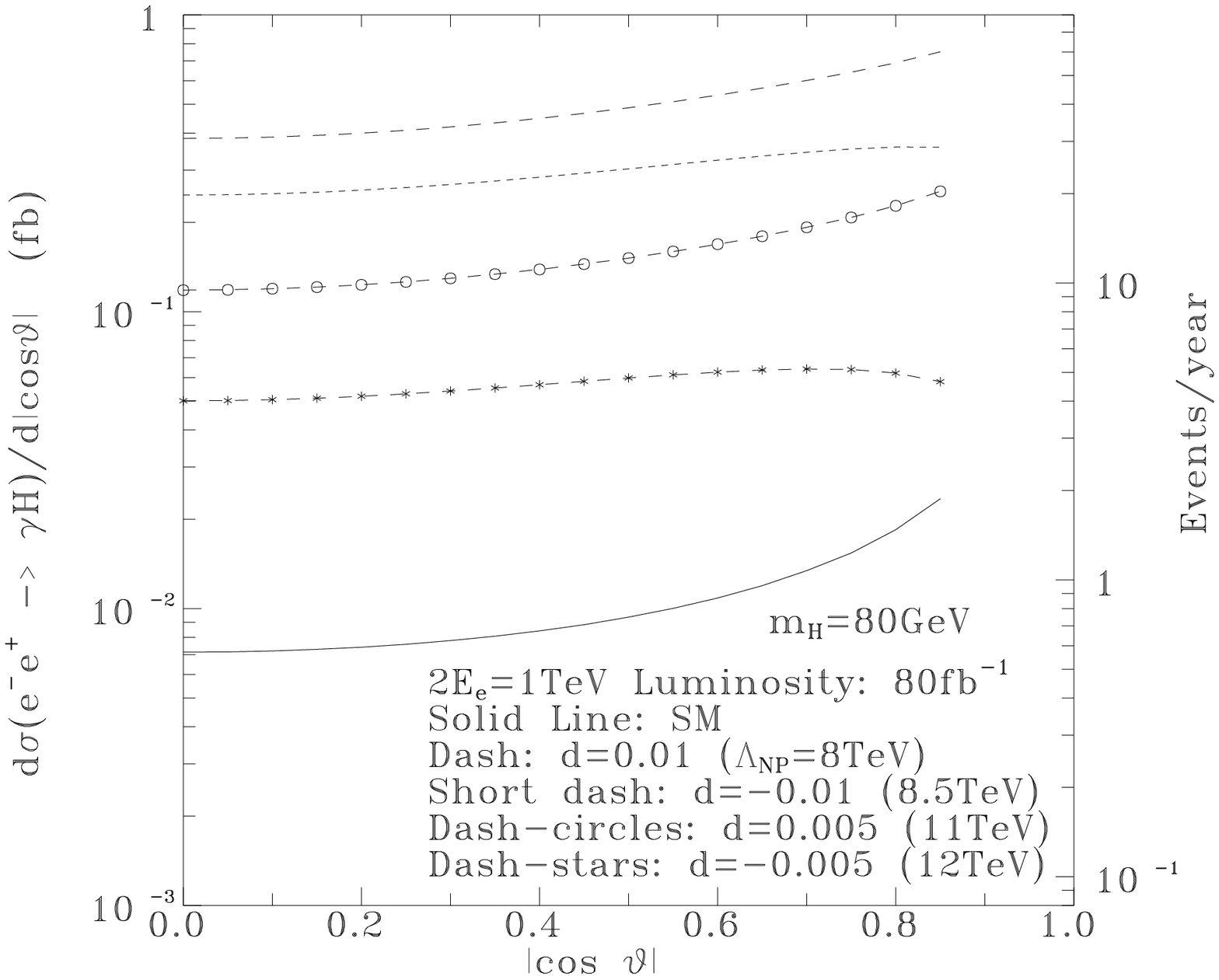}\ \ \ \ \ \ \ \ \
\epsfxsize=7cm
\epsffile[97 0 576 502]{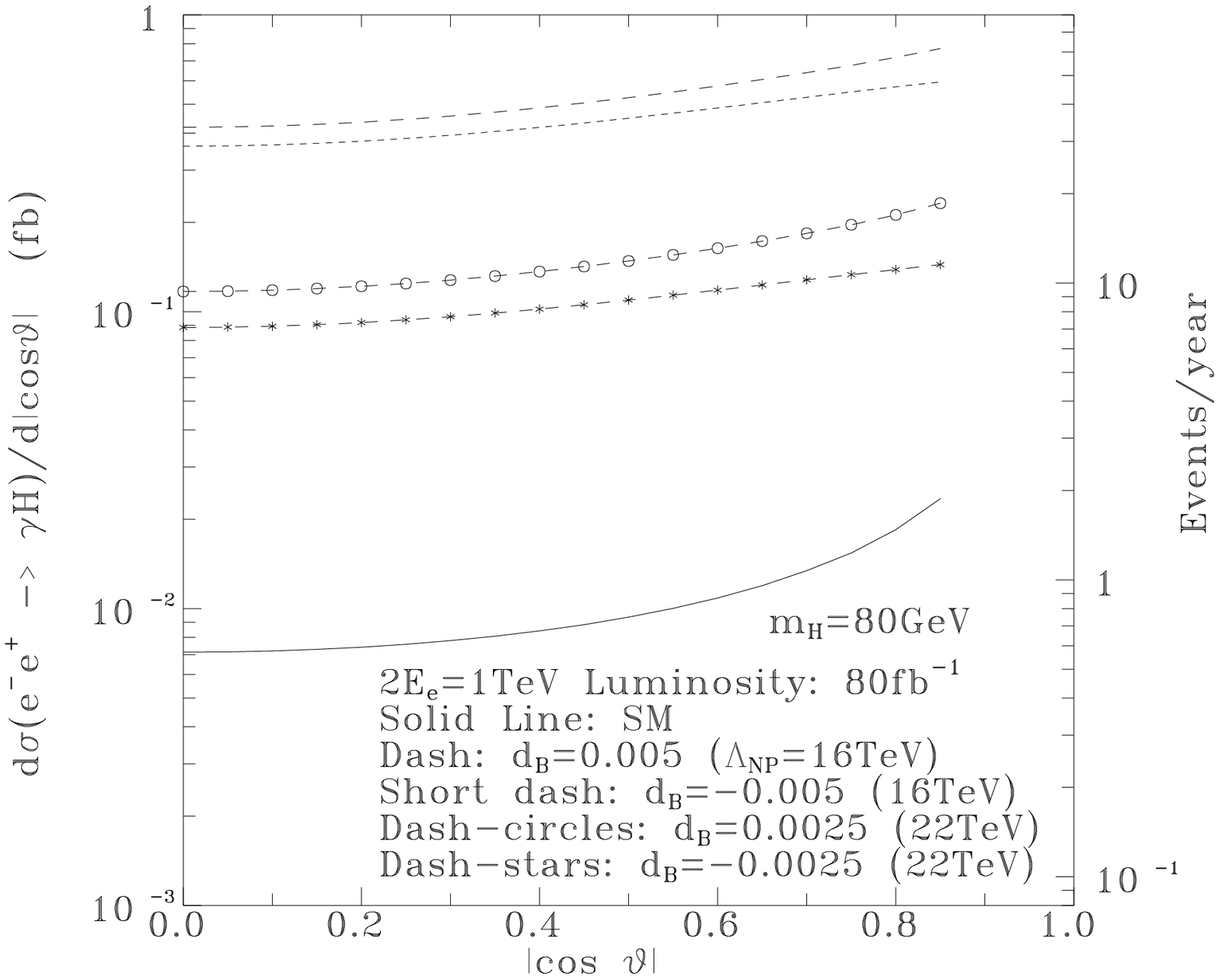}
\]
\vspace{-1.5cm}
\null\\ \vspace{0.5cm}
\centerline{\bf \hfill \  Fig 7a \hfill Fig 7b  \hfill}

\section{Boson Pair Production Processes in $\gamma \gamma$ Collisions}

The boson pair production processes that we consider here
\cite{ggVV2} are
$\gamma \gamma \to W^+W^-$,  $\gamma \gamma \to ZZ$,
$\gamma \gamma \to HH$ and also
$\gamma \gamma \to \gamma Z$,
$\gamma \gamma \to \gamma \gamma$.\par

 An efficient way
of doing this is by looking at the transverse-momentum
distribution of one of the final bosons $B_3$ and $B_4$ and
cutting-off the small $p_T$ values. It is given by
\bq   
{d\sigma \over{dp_Tdy}}~ =~{y p_T\over {8\pi
s_{\gamma\gamma}|\Delta|}}
 \int ^{x_{max}}
_{\tau\over{x_{max}}}\ dxf^{laser}_{\gamma/e}(x)\,
f^{laser}_
{{\gamma /e}}\left({\tau \over{x}}\right) \Sigma|F(\gamma
\gamma \to B_3 B_4|^2 \ ,
\eq
where 
$y \equiv \sqrt{\tau}~ \equiv ~\sqrt { \frac{s_{\gamma
\gamma}}{s_{ee}}}$,
$|\Delta|  =
 {1\over2}\sqrt{s_{\gamma\gamma}(s_{\gamma\gamma}-
4(p^2_T+m^2))}$,
$p^2_T~<~{s_{\gamma\gamma}\over4}-m^2$ 
and $F(\gamma \gamma \to B_3 B_4)$ is the invariant amplitude 
of the subprocess given in \cite{ggVV2}.
 $f^{laser}_{{\gamma /e}}(x)$ is the photon flux distribution. \par
The ${d\sigma \over{dp_T}}$ distribution provides a very
useful way for searching for NP, since it not only takes care of
the events lost along the beam pipe, but also because its
measurement does not require full reconstruction of both final bosons,
at it would be the case for ${d\sigma \over dy}$.
For the illustrations below we choose 
$p_T > p^{min}_T = 0.1 TeV/c$.
The correspondingly expected number of events per year
is obtained by multiplying the above distribution by
the integrated $e^+e^-$ annual
luminosity $\bar{\cal L}_{ee}$ 
taken to be $20, 80, 320 fb^{-1} year^{-1}$ for a
0.5, 1. or 2.TeV collider respectively.\par

We now summarize the properties of each of the five
channels and the way they react to the residual NP
lagrangian.
 SM contributes to  $\gamma\gamma \to W^+W^-$
  at tree level (through $W$ 
exchange diagrams in the $t$, 
$u$ channels involving the $\gamma WW$ vertex and the 
$\gamma\gamma WW$ contact term), and at the 1-loop level to the
 other processes. So the
search for NP effects can be done by precision measurements in the
$ W^+W^-$ channel and by looking for an observable enhancement in the
other channels, see Fig.8-11. The results presented below are
obtained in ref.\cite{ggVV2} using asymptotic expressions for the
anomalous parts of the amplitudes in $\gamma \gamma \to W^+W^-$,  
$\gamma \gamma \to ZZ$,
$\gamma \gamma \to \gamma Z$,
$\gamma \gamma \to \gamma \gamma$, but exact expressions in the
$\gamma \gamma \to HH$ case.\par


\[
\epsfxsize=6cm
\epsffile[97 0 576 502]{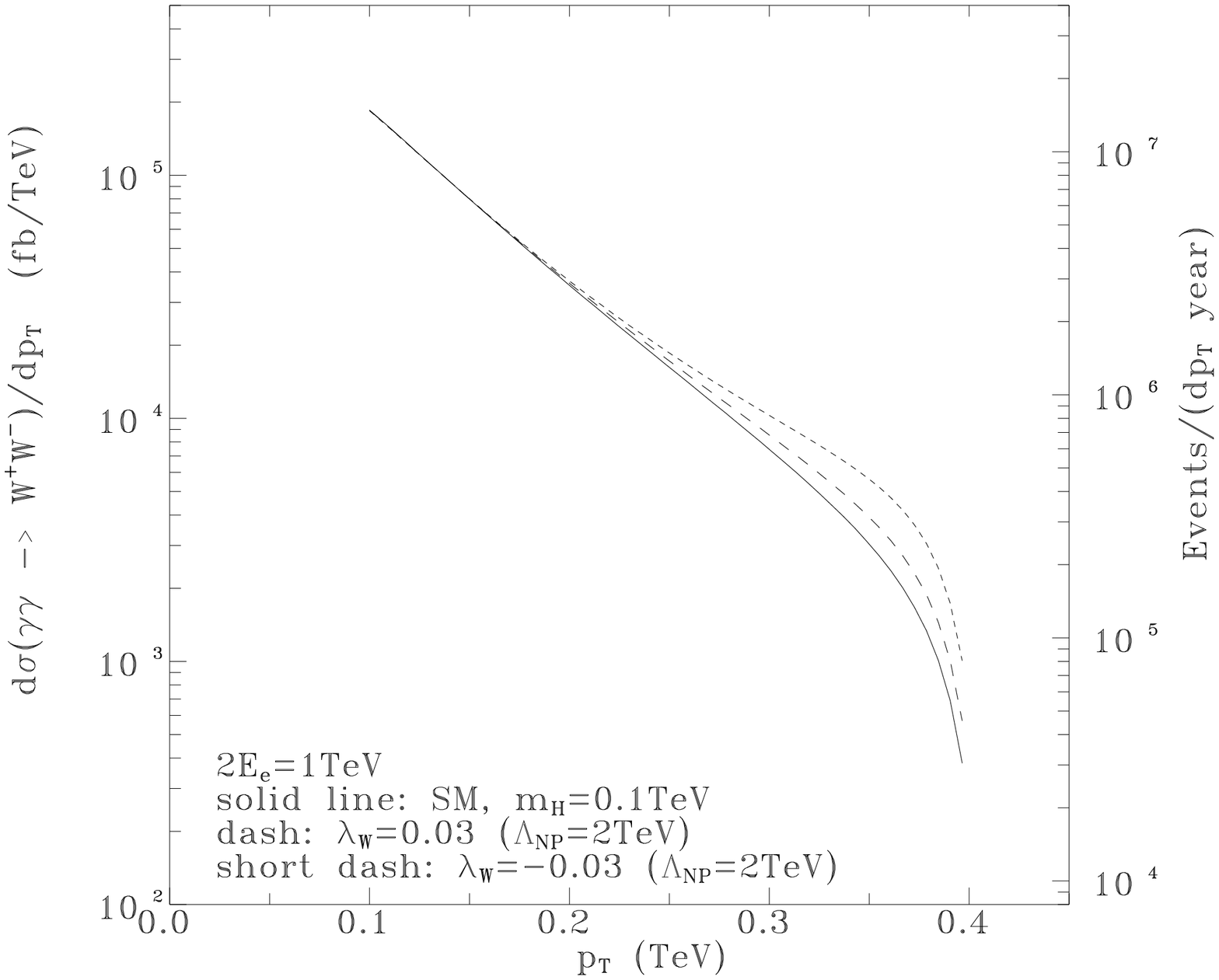}\ \ \ \ \ \ \ \ \ \ \ \ \ \ 
\epsfxsize=6cm
\epsffile[97 0 576 502]{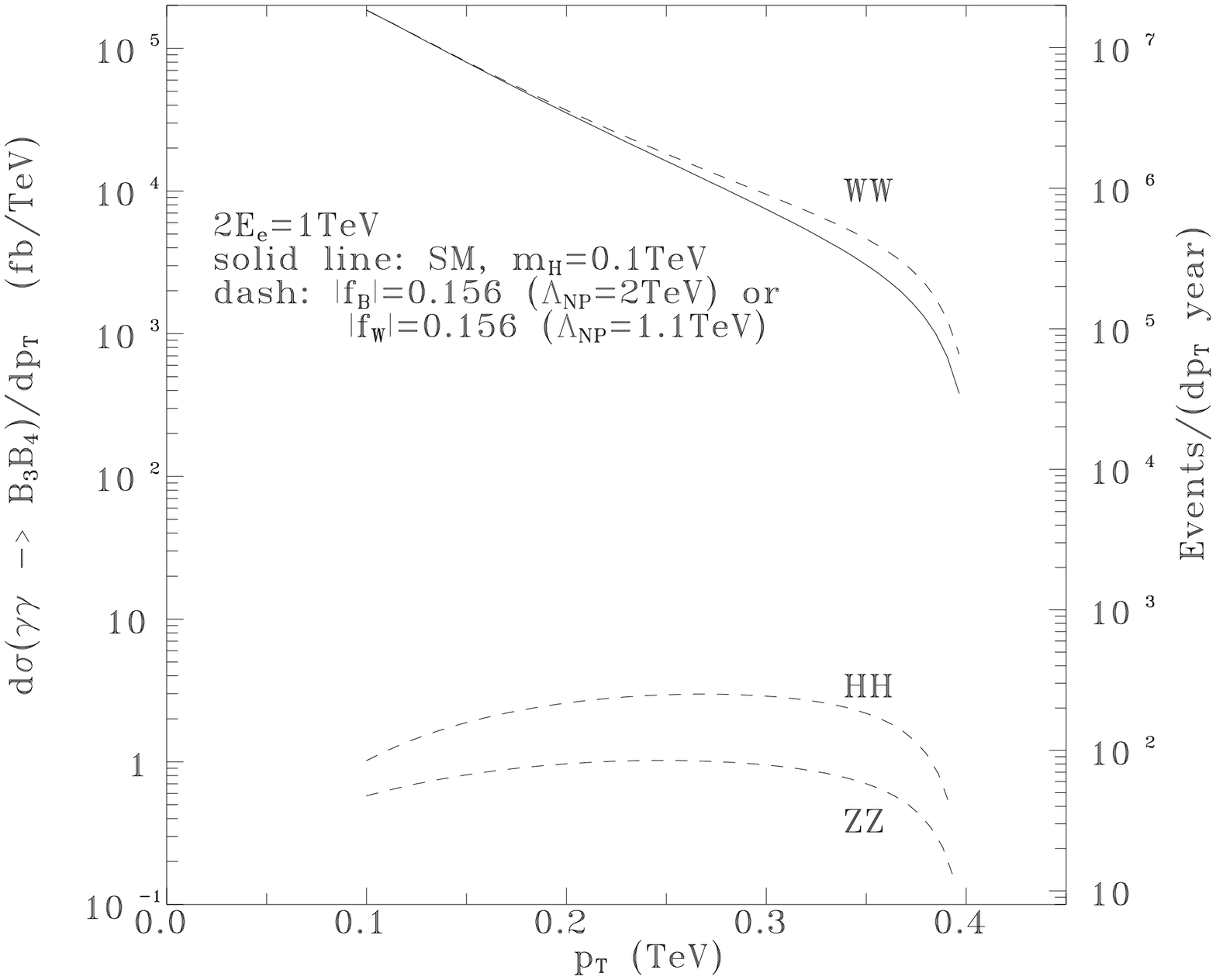}
\]
\vspace{-1.5cm}
\null\\ \vspace{0.5cm}
\centerline{\bf \hfill \  Fig 8 \hfill Fig 9  \hfill}
\[
\epsfxsize=6cm
\epsffile[97 0 576 502]{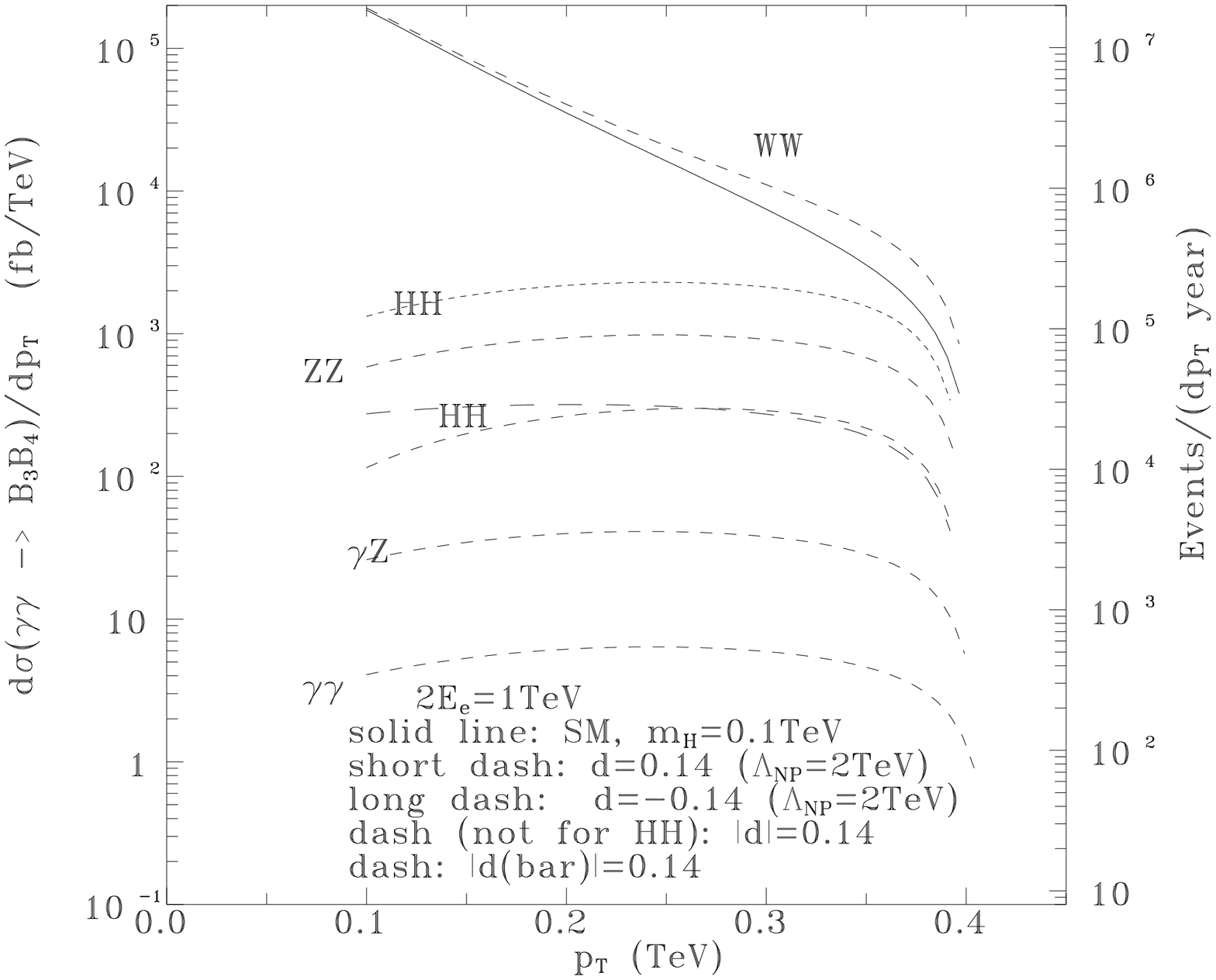}\ \ \ \ \ \ \ \ \ \ \ \ 
\epsfxsize=6cm
\epsffile[97 0 576 502]{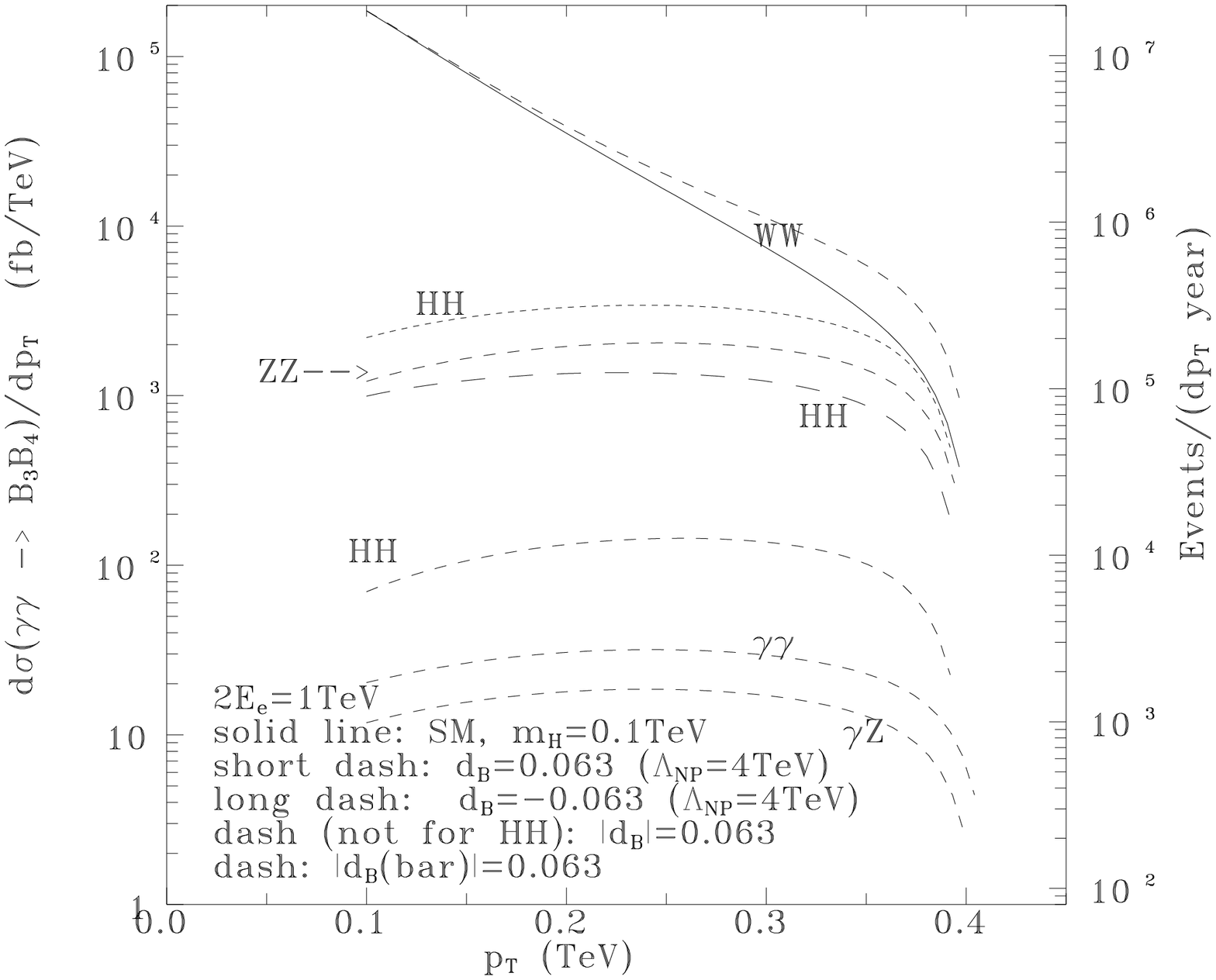}
\]
\vspace{-1.5cm}
\null\\ \vspace{0.5cm}
\centerline{\bf \hfill \  Fig 10 \hfill Fig 11  \hfill}

Table 1 shows how each operator contributes to the various
channels, and one observes that the disentangling is possible between
three groups of operators. In the
$\O_W$ case $W^+W^-$ are produced in (TT) states whereas in the
$\O_{B\Phi}$ and $\O_{W\Phi}$ cases it is mainly (LL). There is no way
to distinguish the contributions of $\O_{B\Phi}$ from that of
$\O_{W\Phi}$ in $\gamma \gamma $ ~collisions. The discrimination
between these two operators requires the use of other processes, like
$e^+e^-\to W^+W^-$. 
The disentangling of
$\O_{UW}$ from $\O_{UB}$ can then be done by looking at the ratios of
the $ZZ$ and $WW$ cross sections. Notice that the linear terms
of the $WW$ and $ZZ$ 
amplitudes for $\O_{UB}$ may be obtained from the corresponding
terms for $\O_{UW}$ by multiplying by the factor  $c^2_W/s^2_W$.

 There is no way to separate the CP-conserving from the
CP-violating terms in these spectra. More detailed spin analyses
are required, like \eg\@ the search for imaginary parts in
final $W$ or $Z$ spin density matrices ~observable through the decay
distributions or the measurement of asymmetries 
associated to linear polarizations of the photon beams.

\renewcommand{\arraystretch}{1.4}
\begin{center}
\begin{tabular}{|c|c|c|cc|cccc|}   \hline
\null &  $SM$ & $\O_{W}$  & $\O_{W\Phi} $ & $\O_{B\Phi}$ & $\O_{UW}$ & 
 $\O_{UB}$  & $\ol\O_{UW}$ & $\ol\O_{UB}$ \\ \hline
 $\gamma\gamma \to WW$ & $TT$ & $TT$ & $LL$ &  $LL$ &
  $LL$ &  $LL$ &  $LL$ & $LL$     \\ \hline
 $\gamma\gamma \to ZZ$ & & & $TT$ & $TT$ &
 $LL$ &  $LL$ &  $LL$ & $LL$     \\ \hline
 $\gamma\gamma \to HH$ &  & &  $X$ & $X$ &
 $X$ &  $X$ &  $X$ &  $X$     \\ \hline
 $\gamma\gamma \to \gamma Z $ &
& & &  &  
 $TT$ &  $TT$ &  $TT$ & $TT$     \\ \hline
 $\gamma\gamma \to \gamma \gamma$ &
& & &  &  
 $TT$ &  $TT$ &  $TT$ & $TT$   \\ \hline  
\end{tabular}
\vspace{0.5cm}\null\\
\centerline{Table 1: Contributions from the various operators}
\end{center}
\noindent

Table 2 summarizes the observability limits expected for each
operator on the basis of $W^+W^-$ production alone. To derive
them we assume that a 5\% departure of the $W^+W^-$ cross
section in the high $p_T$ range from the SM prediction,
will be observable.  These
observability limits on the anomalous couplings give essentially
lower bounds for the couplings to be measurable, that correspond to
upper limits for observable $\Lambda_{NP}$.

\begin{center}
\begin{tabular}{|c|c|c|c|c|c|c|c|c|} \hline
%
 \null &
 \multicolumn{2}{|c|}{ $\O_W$} &
    \multicolumn{2}{|c|}{ $\O_{UW}$ or $\ol{\O}_{UW}$ } &
       \multicolumn{2}{|c|}{ $\O_{UB}$ or $\ol{\O}_{UB}$ } &
    \multicolumn{2}{|c|}{ $\O_{B\Phi}$ or $\O_{W\Phi} $ } \\ \hline 
{$2E_e$(TeV)} &
 {$|\lambda_W|$} &
  {$\Lambda_{NP}$} &
  {$|d|$ or ${|\ol{d}|}$} &
  {$\Lambda_{NP}$} &
   {$|d_B|$ or  $|\ol{d}_B|$ } &
   {$\Lambda_{NP}$} & 
     {$|f_B|$ or $|f_W|$} &
     {$\Lambda_{NP}$}     \\ \hline
 0.5 & 0.04 & 1.7 & 0.1 &2.4  & 0.04 &4.9 & 0.2 & 1.8, 1 \\
 1  & 0.01 &3.5 & 0.04 &4 & 0.015 & 9  & 0.05 & 3.5, 2 \\
 2 & 0.003 & 6.4 & 0.015 & 7 & 0.005 & 16 & 0.015 &6.5, 3.6  \\ \hline
\end{tabular}
\end{center}
\centerline{Table 2: Observability limits based on
the $W^+W^-$ channel}  
\centerline{to the anomalous couplings and the related 
NP scales $\Lambda_{NP}$ (TeV).}
\null\vspace{0.3cm}
\noindent
At 2 TeV, in the case of the operators  $\O_{UB}$, $\O_{UW}$, 
$\ol{\O}_{UB}$, $\ol{\O}_{UW}$ slightly better limits could in
principle be obtained by using the $\gamma\gamma \to ZZ$ process.
Demanding for example that the NP contribution to this process
reaches the level of the SM result for $\gamma\gamma \to ZZ$,  
we can decrease the 
limiting value for the couplings $|d|$
(or $|\ol{d}|$) and $|d_B|$ (or $|\ol{d}_B|$) 
down to 0.01 and 0.003 respectively. This means that a 2 TeV 
collider is sensitive to NP scales up to 20 TeV. 
A more precise analysis using  realistic uncertainties for the 
detection of the ZZ channel and taking into
account the interference between the SM and the NP
contributions,  could probably improve these limits. 
This needs more work.\par

\section{The fusion $\gamma\gamma\to H$}

The cross section for
$\gamma\gamma\to H$ is given by \cite{higpro2}

\bq   
\sigma  = \L_{\gamma \gamma}(\tau_H)~\left({8\pi^2 \over
 m_H} \right)~{\Gamma(H\to\gamma \gamma)\over s_{ee}}
    \ \ \ \ \ \ \ , \ \ \ \ \ \ 
\eq
where the luminosity function $\L_{\gamma\gamma}(\tau_H)$ for 
$\tau_H = {m^2_H / s_{ee}}$ is explicitly given in terms
of the $f^{laser}_{\gamma/e}$ distribution in \cite{higpro2, ggVV2}. 
Essentially $\L_{\gamma\gamma}$ is close to the $e^+e^-$ linear collider
luminosity $\L_{ee}$ up to $\tau_{max}=(0.82)^2$.\par

The $H\to \gamma\gamma$ decay width is computed in terms of the
one-loop SM contribution and tree level NP ones \cite{higpro2} 
obtained from the Lagrangian given in 
eqs. (8):
\bq 
 \Gamma(H\to\gamma \gamma) = {\sqrt2 G_F \over 16\pi} m^3_H
\left (\Big |{\alpha\over4\pi}({4\over3}F_t + F_W) - 2ds^2_W-2d_B c^2_W
\Big |^2  +4|\bar ds^2_W+\bar d_B c^2_W|^2 \right )\ , \ \ \ \ \ 
\eq 
where the standard top ($F_t$) and $W$ ($F_W$)contributions 
are explicitey given in \cite{higpro2}.

The resulting rate is shown in Fig.12-13 for a 1 $TeV$ collider. 
The number of events indicated
in these figures corresponds to an $e^+e^-$ 
luminosity of 80 $fb^{-1}$. 
Strong interference effects may appear (depending
on the Higgs mass) between the SM and the CP-conserving NP 
contributions; (compare Fig.12). But
in the case of CP-violating contributions (Fig.13) there are
never such ~interferences. In the figures we have
only illustrated the cases of $\O_{UW}$ and $\ol{\O}_{UW}$. The
corresponding results for $\O_{UB}$ and $\ol{\O}_{UB}$ can be
deduced according to eq(9) from Fig.12-13, by the replacement $d\,
[\bar d] \to {c^2_W/ s^2_W}d_B\, [\bar d_B]$. 
So the sensitivity to these last two
operators is enhanced by more than a factor 3 in the cross
section.\par


\[
\hspace*{1cm}\epsfxsize=7cm
\epsffile[97 0 576 502]{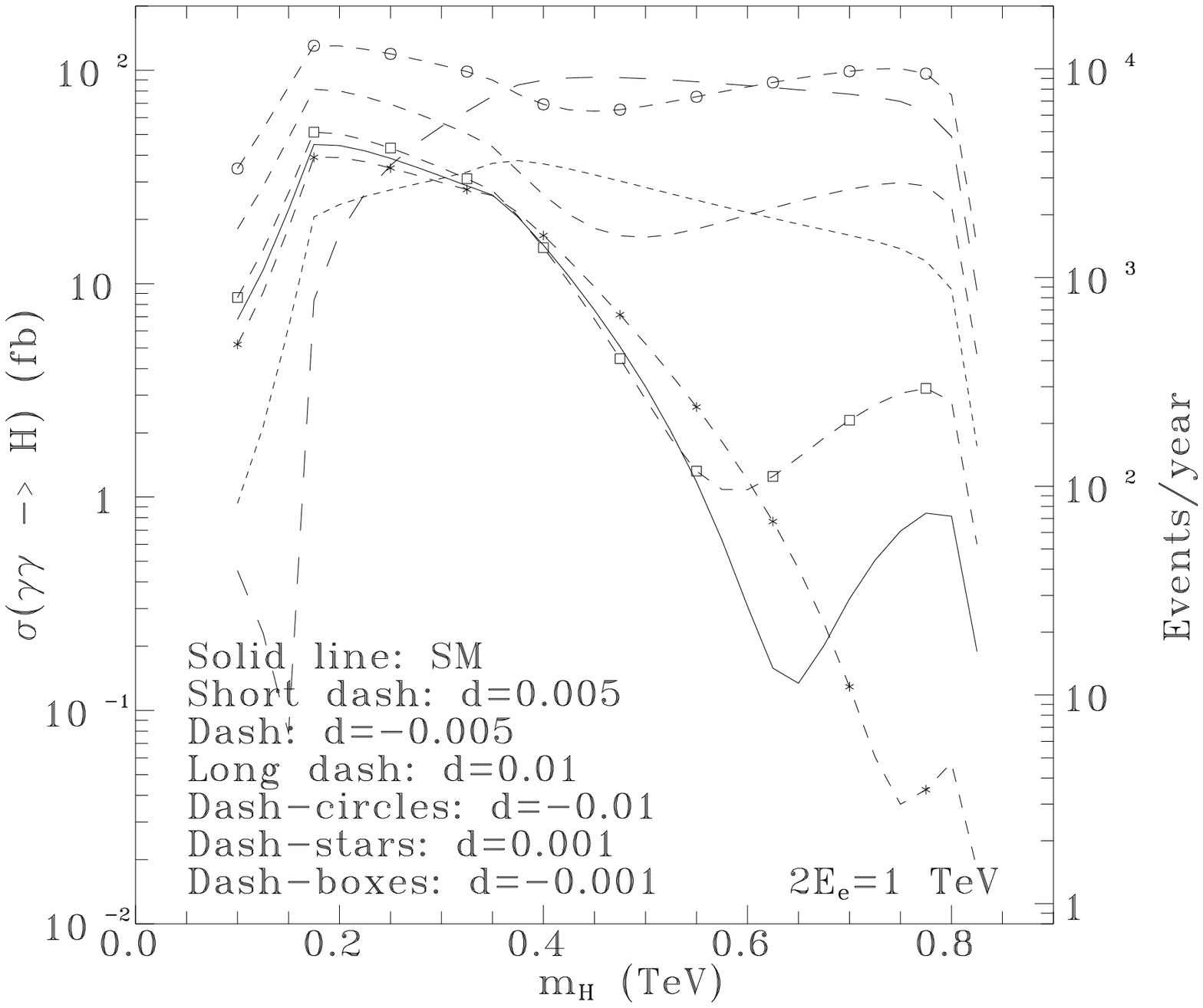}\ \ \ \ \ \ \ \ \
\epsfxsize=7cm
\epsffile[97 0 576 502]{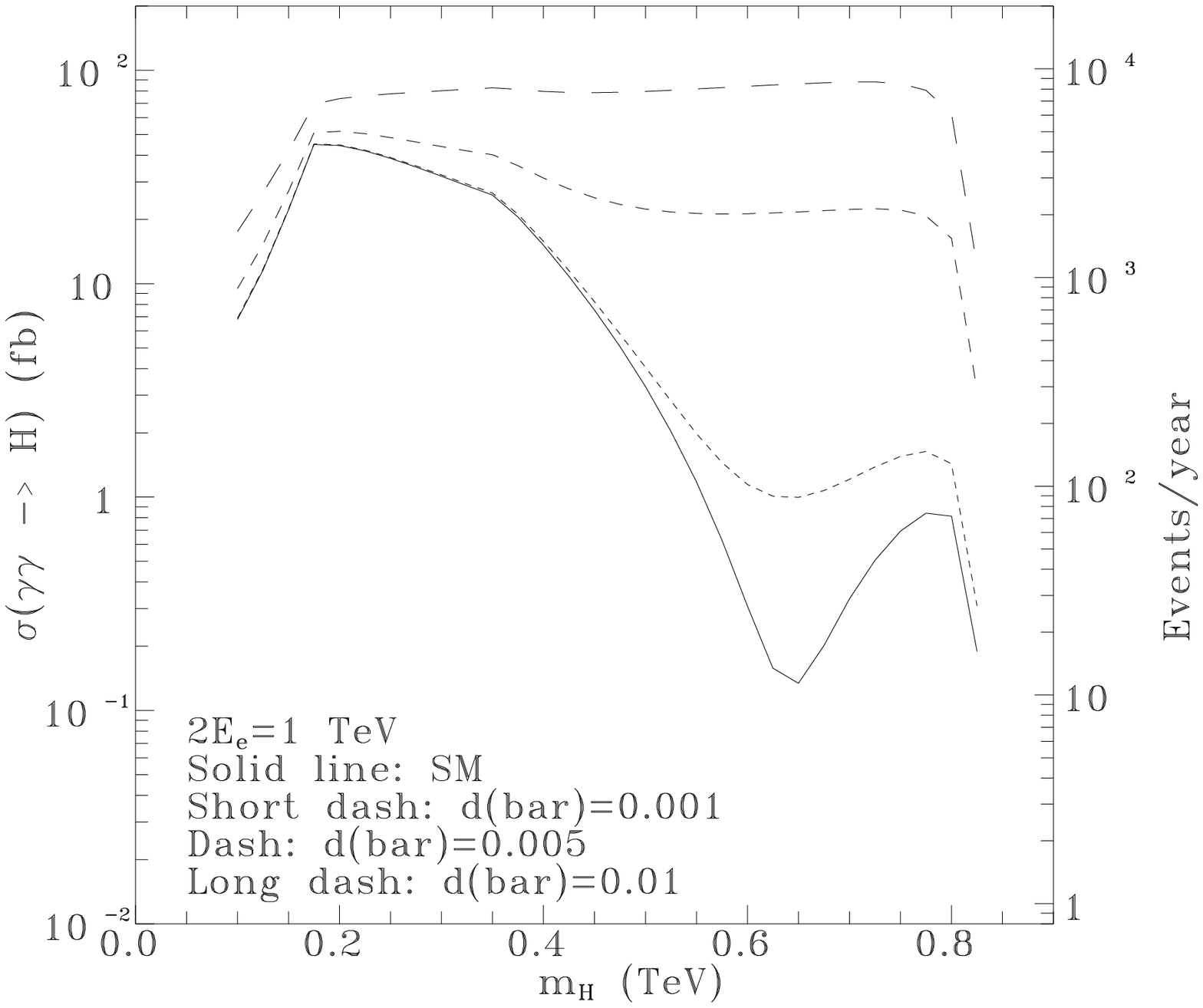}
\]
\vspace{-1.5cm}
\null\\ \vspace{0.5cm}
\centerline{\bf \hfill \  Fig 12 \hfill Fig 13  \hfill}

With the aforementioned designed luminosities, one gets a few thousands
of Higgs bosons produced in the light or intermediate mass range. 
Assuming conservatively an
experimental detection accuracy of about 10\% on the
production rate, one still
gets an observability limit of the order of $10^{-3}$, 
$4.10^{-3}$, $3.10^{-4}$, $10^{-3}$
for $d$, $\bar d$,
$d_B$ and $\bar d_B$ respectively. The corresponding constraints
on the NP scale 
are 26, 13, 65, 36 TeV respectively.\par 

\section{NP searches through ratios of Higgs decay widths}

In this Section we show that independently of the Higgs
production mechanism, the study of the Higgs branching ratios can be
very fruitful for NP searches and for disentangling contributions from
the various operators.

The expression of the $H \to \gamma\gamma$ width has 
been given in (9). Correspondingly, the $H\to\gamma Z$ 
width is also expressed in terms of a
1-loop SM contribution and of the NP contributions from the four
operators.
The $H\to W^+W^-, ZZ$ widths are explicitely given in
 \cite{higpro2}.
They receive a tree level SM contribution
which interferes with the ones from the CP-conserving operators $\O_{UB},
\O_{UW}$. 
Finally we note that the $H\to b\bar b$ decay
width is purely standard and particularly important if
$m_H<140 GeV$.
In Figs. 14a,b the ratios $\Gamma(H\to \gamma\gamma)/\Gamma(H\to b\bar b)$ and
$\Gamma(H\to \gamma Z)/\Gamma(H\to b\bar b)$ are plotted versus $m_H$ 
for a given value of $d$ or $\bar d$. The case of $d_B$ 
or $\bar d_B$ can be obtained by the respective replacements
 $d \to d_B c^2_W / s^2_W$ for the $\gamma\gamma$ amplitude 
and of $d \to d_B$ for the $\gamma Z$ one. The $m_H$ and $d$
dependences  of 
the $\gamma\gamma/b\bar b$ ratio are obviously similar 
to the ones of the production cross section 
$\sigma(\gamma\gamma\to H)$. The ratios $\gamma Z/b\bar b$  
and $\gamma\gamma/\gamma Z$ have independent 
 features that may help disentangling the various couplings. 
They can also be seen from the ratio shown in Fig.14c.\par

\[
\epsfxsize=7cm
\hspace*{1cm}\epsffile[97 0 576 502]{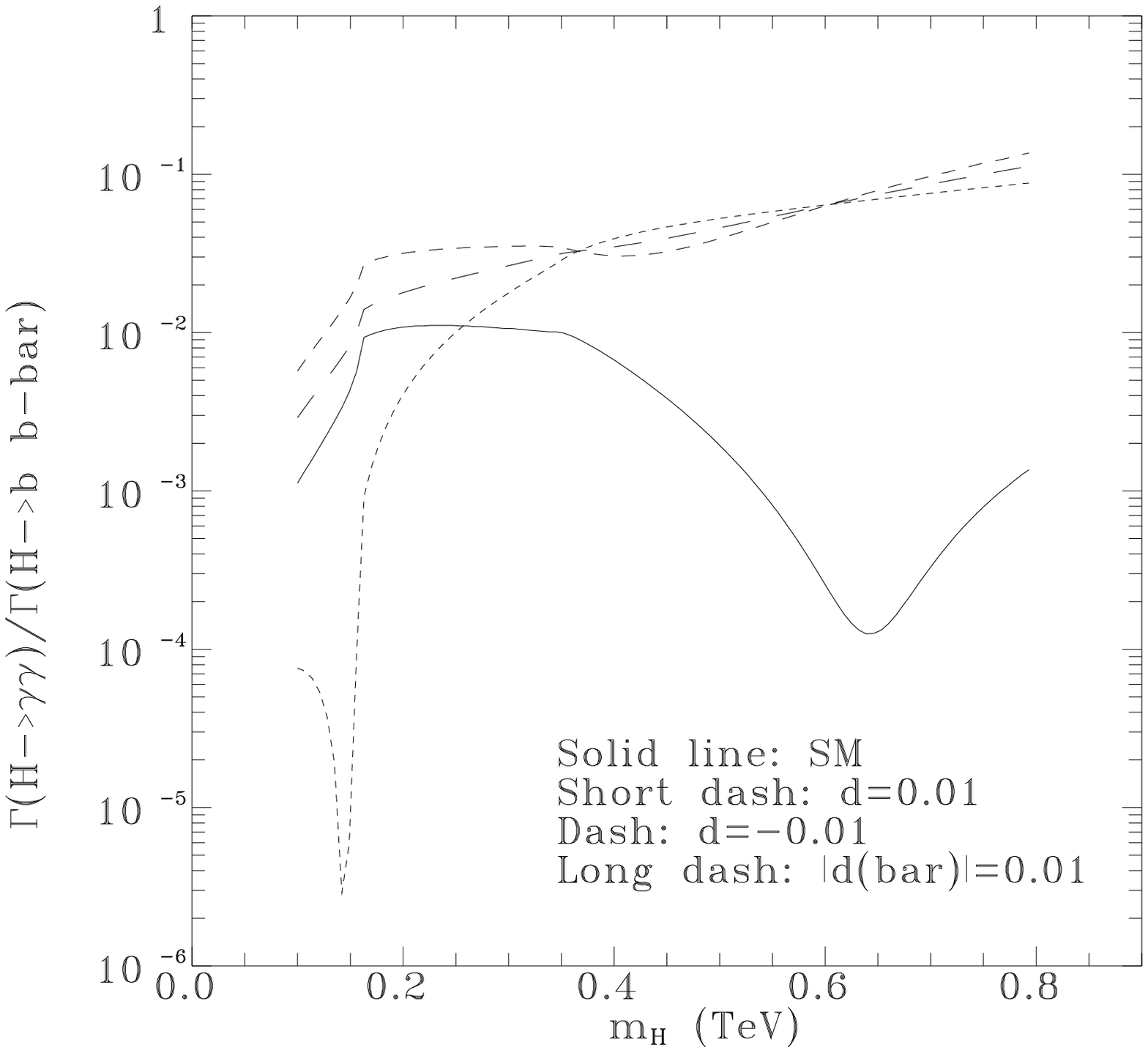}\ \ \ \ \ \ \ \ \
\epsfxsize=7cm
\epsffile[97 0 576 502]{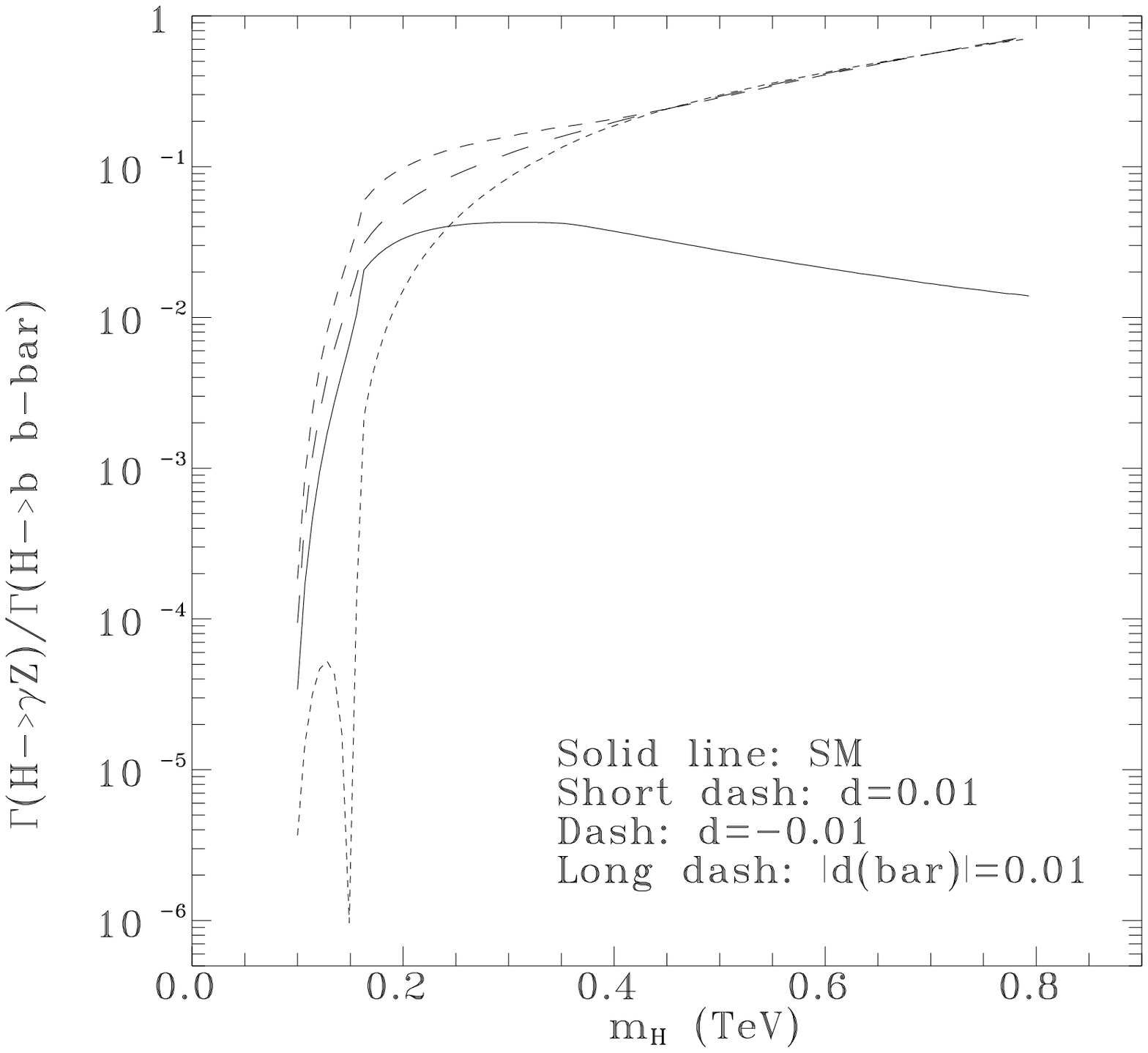}
\]
\vspace{-1.5cm}
\null\\ \vspace{0.5cm}
\centerline{\bf \hfill \  Fig 14a \hfill Fig 14b  \hfill}
\[
\epsfxsize=7cm
\epsffile[97 0 576 502]{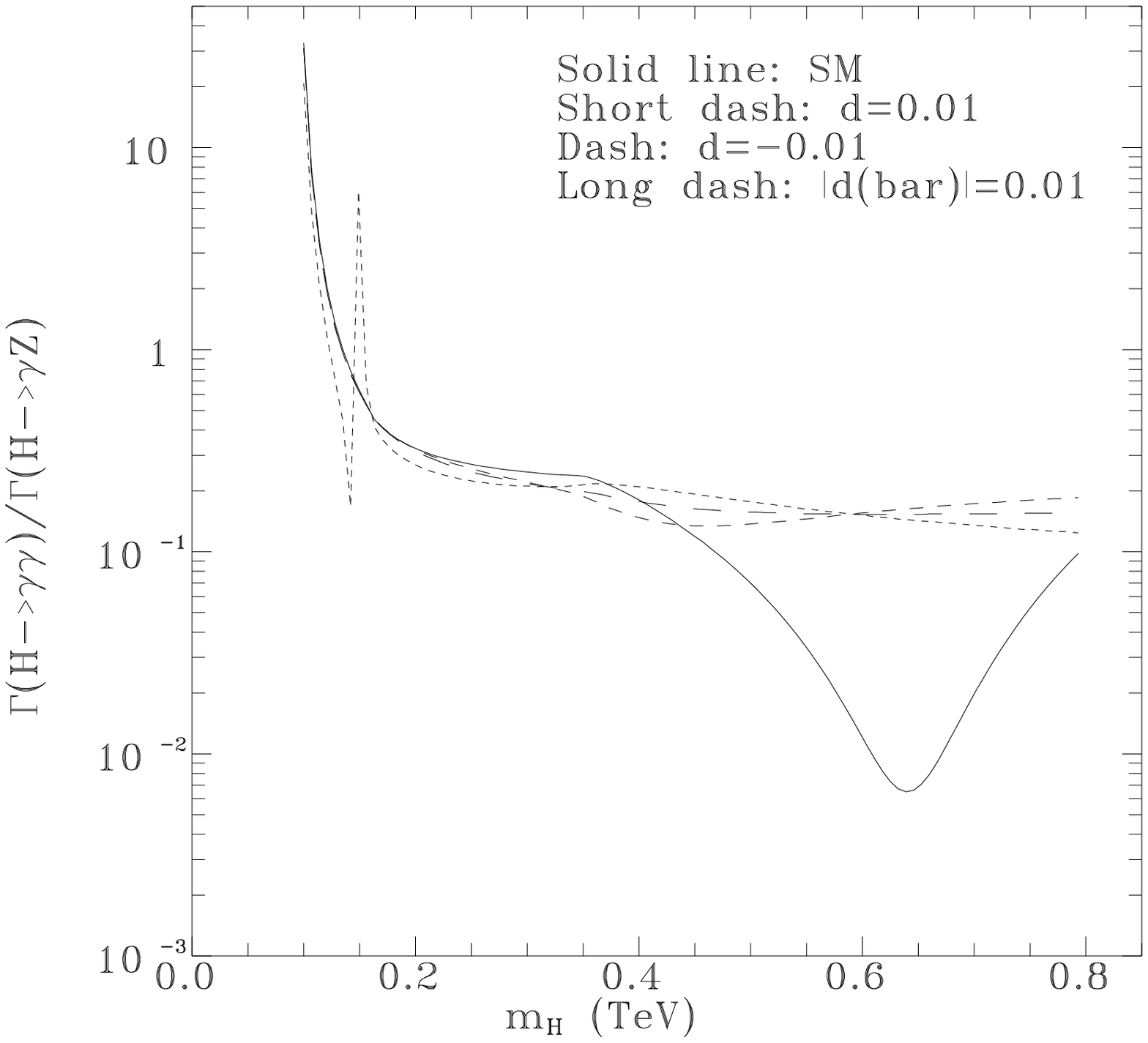} {\bf   Fig 14c}
\]

The sensitivity of the $WW/b\bar b$ and $ZZ/b\bar b$ ratios
is much weaker as expected from the ~occurrence of tree level 
SM contributions. Because of this these ratios are only useful for
$|d| \gsim 0.1$. The ratios $\gamma\gamma/WW$ and $\gamma\gamma/ZZ$
are shown in Figs.14d,e and present
the same features as the ratio $\gamma \gamma/ b\bar b$. 
They can however 
be useful in the range of $m_H$ where the $WW, ZZ$ modes are
dominant.\par

\[
\hspace*{1.5cm}\epsfxsize=7cm
\epsffile[97 0 576 502]{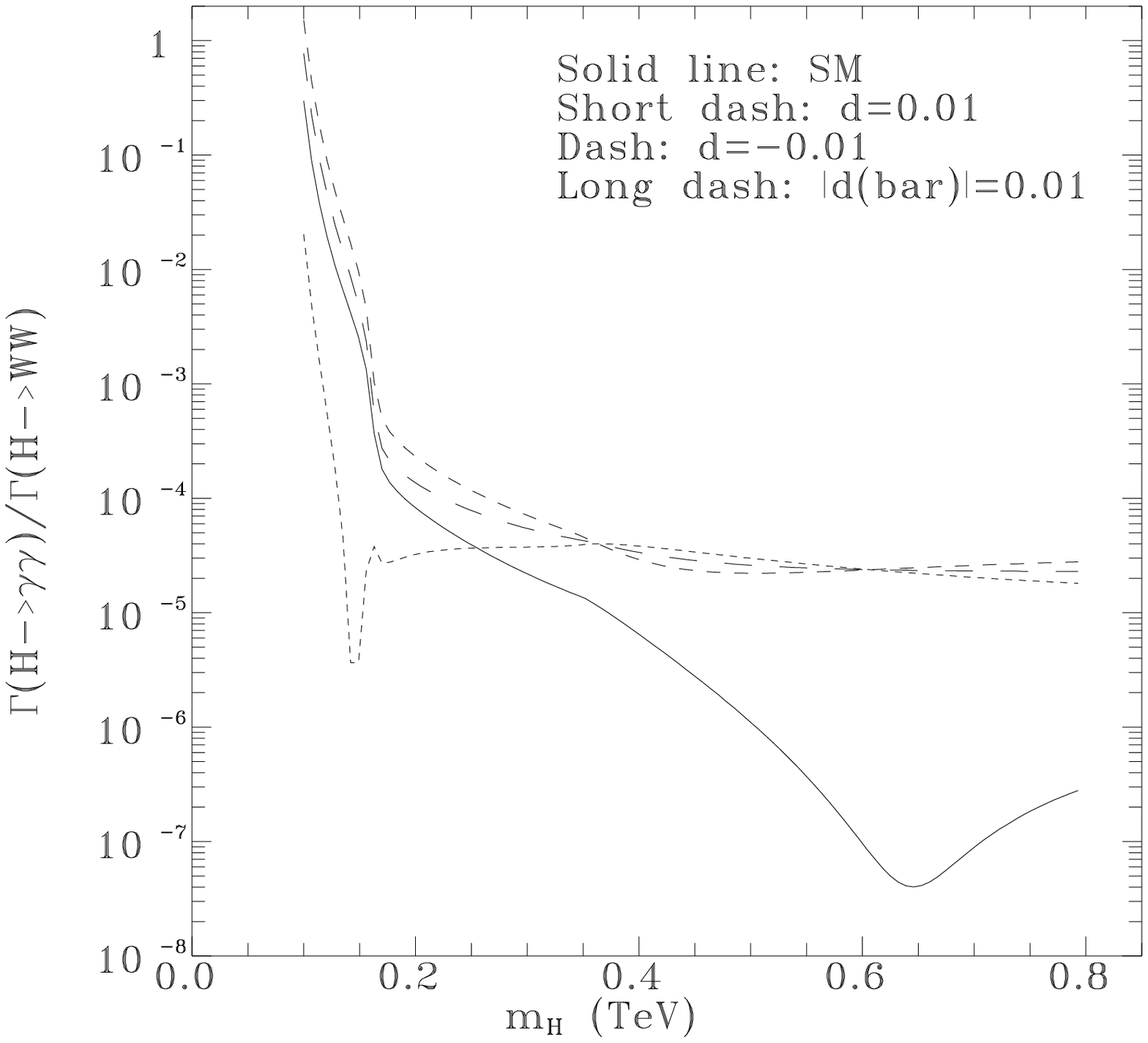}\ \ \ \ \ \ \ \ \
\epsfxsize=7cm
\epsffile[97 0 576 502]{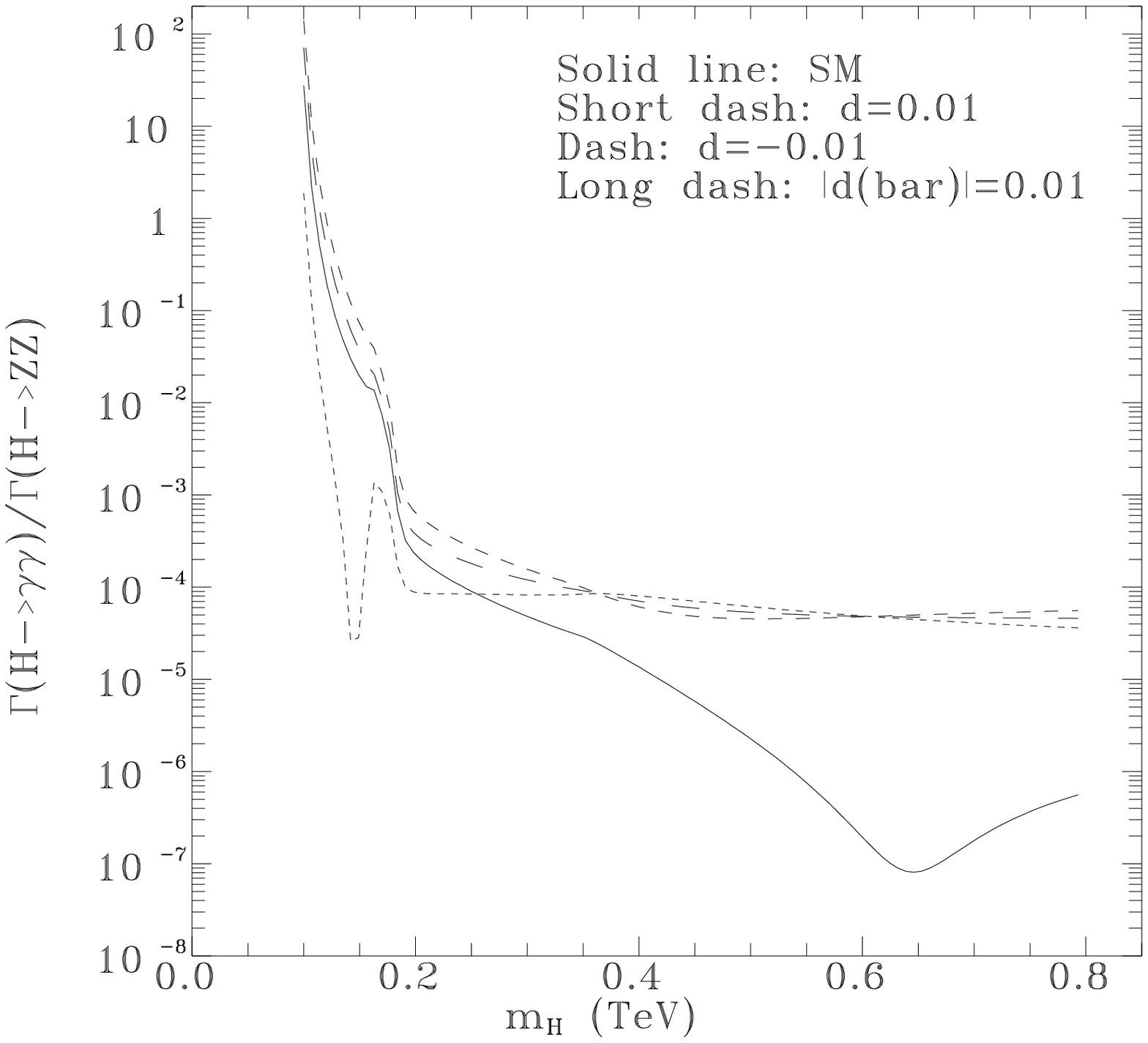}
\]
\vspace{-1.5cm}
\null\\ \vspace{0.5cm}
\centerline{\bf \hfill \  Fig 14d \hfill Fig 14e  \hfill}
%
%
\[
\hspace*{1.5cm}\epsfxsize=7cm
\epsffile[97 0 576 502]{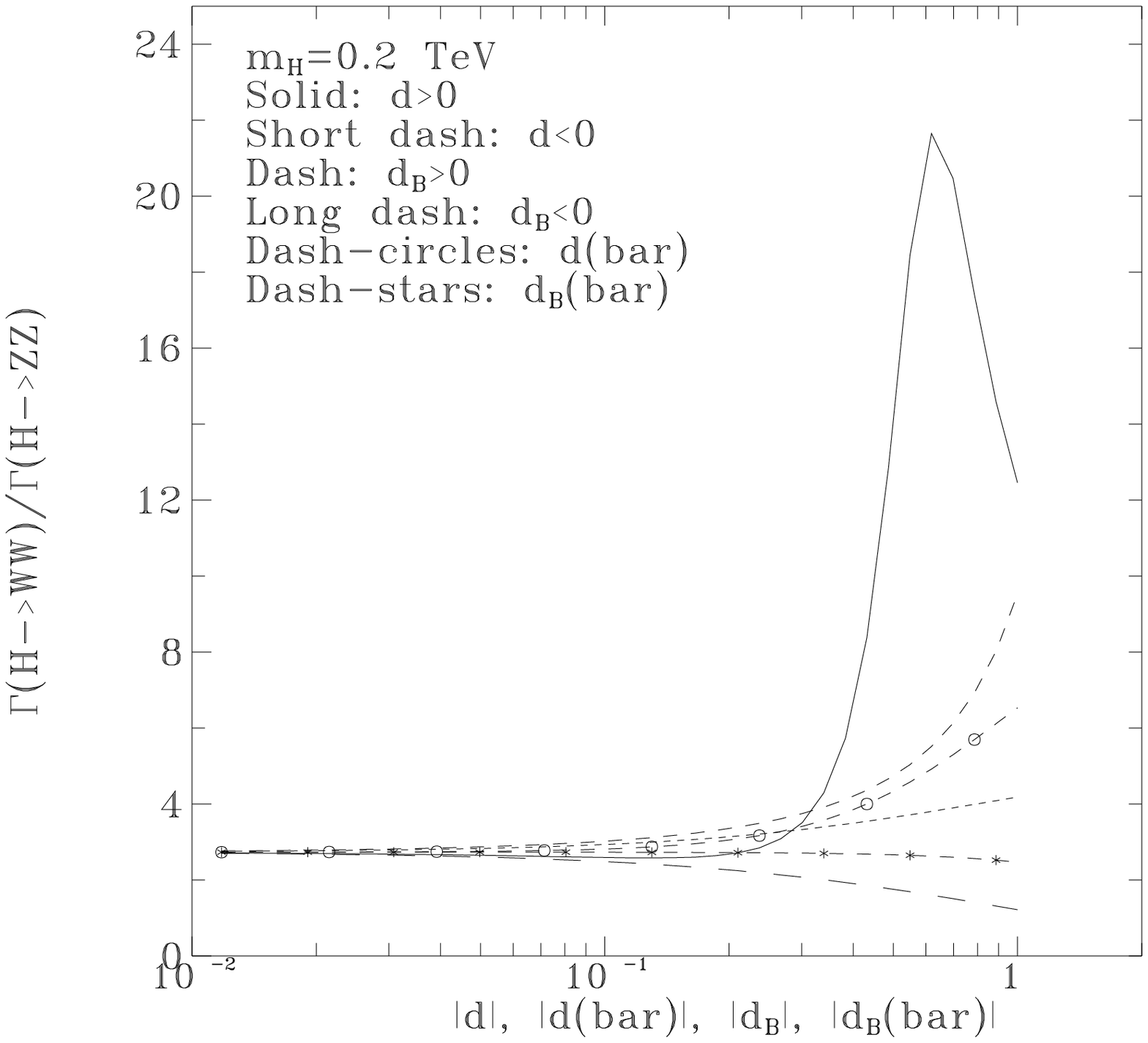}\ \ \ \ \ \ \ \ \
\epsfxsize=7cm
\epsffile[97 0 576 502]{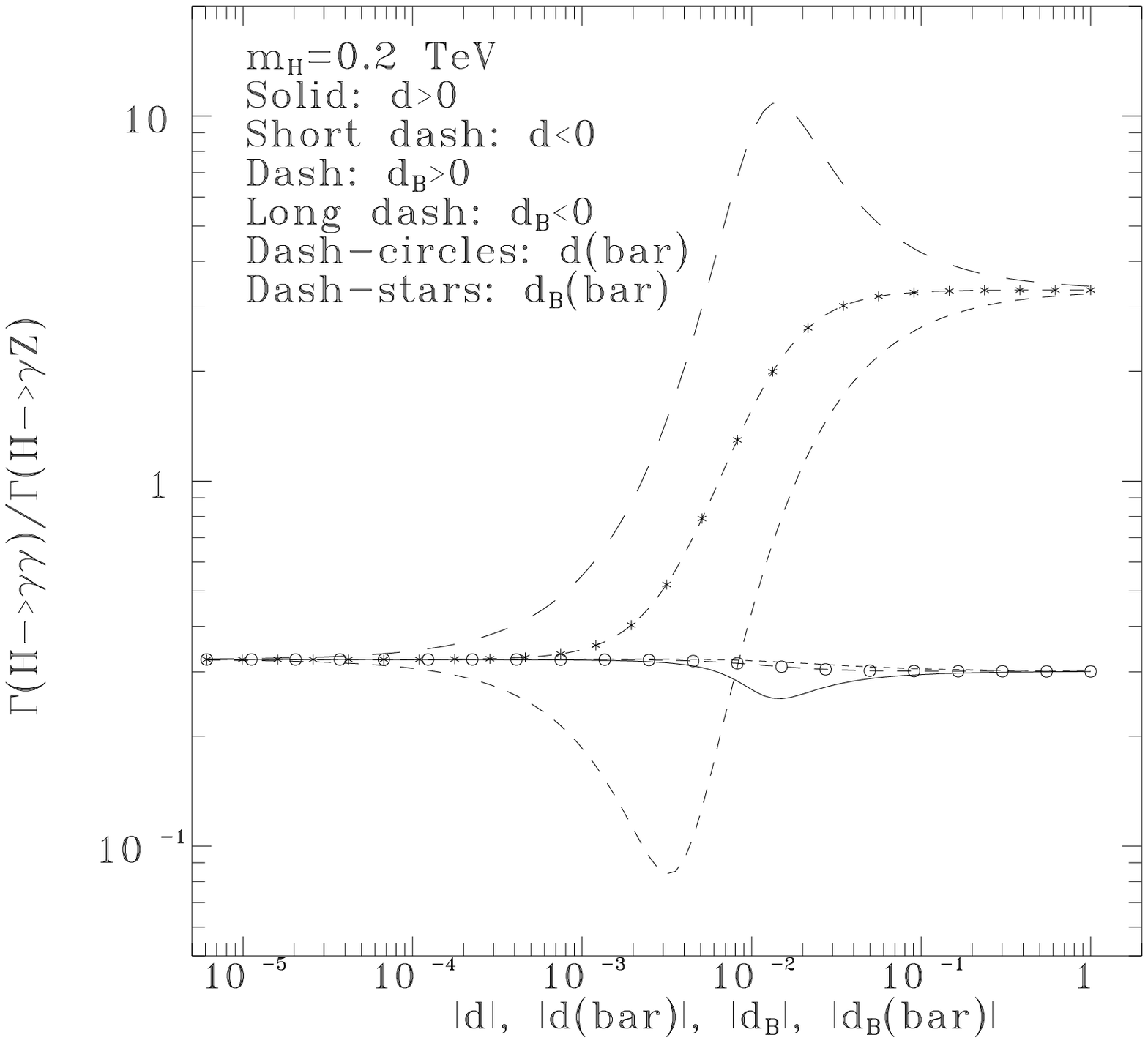}
\]
\vspace{-1.5cm}
\null\\ \vspace{0.5cm}
\centerline{\bf \hfill \  Fig 15a \hfill Fig 15b  \hfill}
Finally we discuss the ratios $WW/ZZ$ and $\gamma\gamma/\gamma
Z$ for a given value of $m_H$ (chosen as $0.2 TeV$ in 
the illustrations made in Fig.15a,b), 
versus the coupling constant values of
the four operators. This shows very ~explicitly how these
ratios can be used for disentangling the various operators.
They have to be taken in a complementary way to the 
$\sigma(\gamma\gamma\to H)$ measurement. 
As in the $WW/b \bar b$ case, the $WW/ZZ$ ratio is only useful
for $|d| \gsim 0.1$, while $\gamma\gamma/\gamma Z$
ratio allows for disentangling $d$ values down to $10^{-3}$
or even less. The corresponding sensitivity limit for $d_B,\,
\bar d_B$ should then be lying at the level of a few times
$10^{-4}$.
\section{Final discussion}

We have concentrated our analysis on the residual NP effects
described
by seven blind and one superblind $dim=6$ $SU(2)\times U(1)$ gauge 
invariant operators $\O_W$, $\O_{B\Phi}$, $\O_{W\Phi}$, 
${\cal O}_{UB}$, 
${\cal O}_{UW}$, $\overline{{\cal O}}_{UB}$, $\overline{{\cal
O}}_{UW}$, ${\cal O}_{\Phi 2}$. The first three ones should be
essentially studied in the process $e^+e^- \to W^+W^-$ but they
also contribute to $\gamma\gamma \to WW$. The five other ones
only affect the Higgs
couplings to themselves and to
gauge bosons, and their effects have been considered in all processes
in which a Higgs can be exchanged or produced.\par

We started with the Higgs production process 
$e^{-}e^{+}\to ZH$. 
We have analyzed the angular distributions for $HZ$
as well as the four azimuthal asymmetries in $Z\to f\bar
f$ decays associated to the $\cos
\phi _f$, $\sin \phi _f$, $\sin 2\phi _f$ and $\cos 2\phi _f$ 
distributions controlled by the
$Z$ spin density matrix elements. The 
disentangling of the five relevant operators 
should be achievable down to the percent level for the NP
couplings, implying sensitivities to NP scales of the
order of 10, 8 and 12 $TeV$ for ${\cal O}_{\Phi 2}$, ${\cal
O}_{UB}$ and ${\cal O}_{UW}$ respectively.
This study is further augmented by looking at 
 the angular distribution of $e^{-}e^{+}\to \gamma \gamma $,
which is
sensitive to a different combination of the
${\cal O}_{UB}$ and ${\cal O}_{UW}$ couplings, implying
also sensitivities to NP scales of 20
and 12 $TeV$.

We have then considered $\gamma\gamma $ collision processes.
In boson pair production one can feel 
NP effects associated to scales up to 
$\Lambda_{NP}\simeq 20 TeV$  for collider energies up to $2TeV$. 
This can be achieved by simply measuring final
gauge boson $p_T$ distributions.
A comparison of the effects in the various final
states $WW$, $ZZ$, $HH$, $\gamma Z$ and  $\gamma\gamma$ including
an analysis of final spin states,
(i.e. separating $W_T(Z_T)$ from $W_L(Z_L)$ states),
should
allow a selection among the seven candidate operators.\par

At a linear $e^+e^-$ collider the cross section of the
process $\gamma\gamma\to H$
is the best way to study the Higgs boson, if its mass allows
its production.
We showed that the sensitivity limits on the NP couplings are at
the $10^{-3}$ level, implying new physics scales 
up to 65 TeV, depending on the nature of the
NP operator. \par

We have then shown that independently of the production mechanism,
having at our disposal a few thousands of Higgs produced (either in
$e^+e^-\to HZ$ or in $\gamma\gamma\to H$), enables the  
disentangling of the 
 $\O_{UB}$ 
and $\O_{UW}$ operators by looking at the Higgs
branching ratios into $WW$, $ZZ$, $\gamma\gamma$, $\gamma Z$.
Because these branching ratios  
react  differently to the presence of each of these operators.
This is illustrated with the ratios of these channels to the $b\bar b$
one which is unaffected by this kind of NP.
Most spectacular for the disentangling of the various operators
seem to be the ratios $WW/ZZ$ and $\gamma\gamma/\gamma Z$. 
The first one, which is applicable in the intermediate
and high Higgs mass range, allows to disentangle $\O_{UB}$ 
from $\O_{UW}$ down to values of the order of $10^{-1}$, whereas the
second one, applicable in the light Higgs case, is sensitive to
couplings down to the $10^{-3}$ level or less.\par

A direct identification of CP
violation requires
either an analysis of the W or Z spin density matrix through
their fermionic decay distributions \cite{hzVlachos}, or
the observation of a suitable asymmetry with linearly polarized 
photon beams.

Finally we emphasize that the ~occurrence of anomalous terms in 
gauge or Higgs boson couplings would be of great interest for tracing
 the origin of NP and its basic properties.\par

{\bf \underline{Acknowledgements}}\par
This work has been partially supported by the EC contract
CHRX-CT94-0579.

\end{document}